\documentclass[3p,english,12pt,times,sort&compress,nopreprintline]{elsarticle}
\usepackage{fullpage}

\usepackage[utf8]{inputenc}
\usepackage[T1]{fontenc}
\usepackage{babel}
\usepackage{array}
\usepackage{multirow}
\usepackage{amsmath}
\usepackage{amsthm}
\usepackage{amssymb}
\usepackage{nicefrac}
\usepackage{graphicx}
\usepackage[unicode=true]{hyperref}
\usepackage{xcolor}
\usepackage{textgreek}
\usepackage{comment}
\usepackage[title]{appendix}
\usepackage{subfig}
\usepackage{longtable}
\usepackage{multirow}
\usepackage{booktabs}
\usepackage{caption}
\usepackage{array}
\usepackage{calc}
\usepackage{etoolbox}

\hypersetup{
  colorlinks=true,
  linkcolor=blue,
  filecolor=magenta,
  urlcolor=cyan,
}

\makeatletter
\newcommand{\orcidID}[1]{\href{https://orcid.org/#1}{\includegraphics[width=10pt]{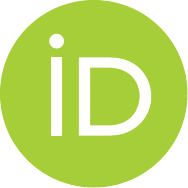}}}
\def\autocite#1\citep{#1}
\theoremstyle{plain}

\newtheorem{theorem}{\protect\theoremname}
\newtheorem*{theorem*}{\protect\theoremname}
\theoremstyle{plain}

\newtheorem{definition}[theorem]{\protect\definitionname}
\newtheorem*{definition*}{\protect\definitionname}
\theoremstyle{plain}

\newtheorem*{remark*}{\protect\remarkname}
\theoremstyle{plain}

\newtheorem{lemma}[theorem]{\protect\lemmaname}
\newtheorem*{lemma*}{\protect\lemmaname}
\theoremstyle{plain}
\newtheorem{corollary}[theorem]{\protect\corollaryname}
\newtheorem*{corollary*}{\protect\corollaryname}
\makeatother

\providecommand{\definitionname}{Definition}
\providecommand{\lemmaname}{Lemma}
\providecommand{\remarkname}{Remark}
\providecommand{\theoremname}{Theorem}
\providecommand{\corollaryname}{Corollary}

\usepackage{aliascnt}
\makeatletter
\let\corollary\@undefined
\let\endcorollary\@undefined
\let\lemma\@undefined
\let\endlemma\@undefined
\let\definition\@undefined
\let\enddefinition\@undefined
\makeatother
\theoremstyle{plain}
\newaliascnt{lemma}{theorem}
\newtheorem{lemma}[lemma]{Lemma}
\aliascntresetthe{lemma}

\newaliascnt{corollary}{theorem}
\newtheorem{corollary}[corollary]{Corollary}
\aliascntresetthe{corollary}

\newaliascnt{definition}{theorem}
\newtheorem{definition}[definition]{Definition}
\aliascntresetthe{definition}

\makeatletter
\newcommand\Autoref[1]{\@first@ref#1,@}
\def\@throw@dot#1.#2@{#1}
\def\@set@refname#1{
    \edef\@tmp{\getrefbykeydefault{#1}{anchor}{}}%
    \xdef\@tmp{\expandafter\@throw@dot\@tmp.@}%
    \ltx@IfUndefined{\@tmp autorefnameplural}%
         {\def\@refname{\@nameuse{\@tmp autorefname}s}}%
         {\def\@refname{\@nameuse{\@tmp autorefnameplural}}}%
}
\def\@first@ref#1,#2{%
  \ifx#2@\autoref{#1}\let\@nextref\@gobble
  \else%
    \@set@refname{#1}
    \@refname~\ref{#1}
    \let\@nextref\@next@ref
  \fi%
  \@nextref#2%
}
\def\@next@ref#1,#2{%
   \ifx#2@ and~\ref{#1}\let\@nextref\@gobble
   \else, \ref{#1}
   \fi%
   \@nextref#2%
}
\makeatother

\makeatletter
\def\maxwidth{\ifdim\Gin@nat@width>\linewidth\linewidth\else\Gin@nat@width\fi}
\def\maxheight{\ifdim\Gin@nat@height>\textheight\textheight\else\Gin@nat@height\fi}
\makeatother
\setkeys{Gin}{width=\maxwidth,height=\maxheight,keepaspectratio}

\begin{document}

\begin{frontmatter}

\title{The Complex-Pole Filter Representation (COFRE) for spectral modeling of fNIRS signals}

\author[mek]{Marco A. Pinto Orellana \orcidID{0000-0001-6495-1305}}
\corref{cor}
\author[mek]{Peyman Mirtaheri \orcidID{0000-0002-7664-5513}}
\author[it,simulamet]{Hugo L. Hammer \orcidID{0000-0001-9429-7148}}

\cortext[cor]{Corresponding author}

\address[mek]{Department of Mechanical, Electronics and Chemical Engineering. Oslo Metropolitan University.}
\address[it]{Department of Information Technology. Oslo Metropolitan University.}
\address[simulamet]{Department of Holistic Systems, Simula Metropolitan Center for Digital Engineering.}

\begin{abstract}
The complex-pole frequency representation (COFRE) is introduced in this paper as a new approach for spectrum modeling in biomedical signals. Our method allows us to estimate the spectral power density at precise frequencies using an array of narrow band-pass filters with single complex poles. Closed-form expressions for the frequency resolution and transient time response of the proposed filters have also been formulated. In addition, COFRE filters have a constant time and space complexity allowing their use in real-time environments. Our model was applied to identify frequency markers that characterize tinnitus in very-low-frequency oscillations within functional near-infrared spectroscopy (fNIRS) signals. We examined data from six patients with subjective tinnitus and seven healthy participants as a control group. A significant decrease in the spectrum power was observed in tinnitus patients in the left temporal lobe. In particular, we identified several tinnitus signatures in the spectral hemodynamic information, including (a.) a significant spectrum difference in one specific harmonic in the metabolic/endothelial frequency region, at 7mHz, for both chromophores and hemispheres; and (b.) a significant differences in the range 30-50mHz in the neurogenic/myogenic band.
\end{abstract}


\end{frontmatter}

\section{Introduction}

Tinnitus is an unintentional experience of meaningless sounds in the absence of external sources \citetext{\citealp[p.~230]{TextbookTinnitus-Moller-2011}; \citealp[p.~121]{TinnitusTreatmentClinical-Tyler-2006}} and it can be primarily affected by environmental and physiological factors \citep{NeuroPhysApproachTinnitus-Jastreboff-1996, PhantomAuditoryPerception-Jastreboff-1990, TinnitusMultidisciplinaryApproach-Baguley-2013}. This type of auditory hallucination provokes neural oscillation patterns in the auditory cortex that resemble waveforms generated due to sound stimuli \citep[p.~121-122]{TinnitusTreatmentClinical-Tyler-2006}. We contribute to analyzing this health condition by providing a method for extracting accurate hemodynamic information in the spectrum domain.

Tinnitus involves a rise in spontaneous firing and synchronization of the neuronal activity in the auditory cortex. These changes induce variations in the spectral properties of the patient's brain hemodynamic and electrical signals. For instance, changes in the spectrum power in brain signals play an essential function in predicting of a chronic tinnitus level \citep[p.~163]{TextbookTinnitus-Moller-2011}. In addition, tinnitus-provoked increments of the synaptic metabolism lead to abnormal transient responses in the oxy-hemoglobin concentration \citep[p.~57]{NeuroscienceTinnitus-Eggermont-2012}. Furthermore, drastic syncope-inducing shifts of low blood pressure on the brain, which involve sudden hemodynamic imbalances, can also generate episodes of tinnitus \citep[p.~88]{TextbookTinnitus-Moller-2011}.

These tinnitus-associated hemodynamic changes have often been studied using functional near-infrared spectroscopy (fNIRS) as an imaging method. fNIRS is a non-invasive imaging technique for measuring concentration changes of chromophores, oxy-hemoglobin (HbO) and deoxy-hemoglobin (HbR), using light in the near-infrared region (700-900 nm). In a conventional continuous fNIRS device, two light sources separated by 2--3 cm with different optical wavelengths emit light to the brain. Chromophore concentrations are then estimated from the received light intensity using the modified Beer-Lambert law. Research from Sevy et al. \citep{NeuroimagingNIR-Sevy-2010} and Olds et al. \citep{CorticalActivationPatterns-Olds-2016} validated fNIRS to detect tinnitus-cause hemodynamic variations during cortical stimulation in patients with severe auditory deficits with cochlear implants. Later experiments used fNIRS to illustrate that tinnitus patients exhibit certain unique phenomena in the temporal region due to their anatomical closeness to the auditory cortex: (a.) higher HbO concentrations in the left hemisphere compared with a healthy-subject baseline \citep{FNIR-Schecklmann-2014}; (b.) general activation in both hemispheres \citep{FNIR-Schecklmann-2014}; and (c.) activation steadiness even during quiet periods while activation reduction is visible in healthy subjects \citep{HumanAuditoryAdjacent-Issa-2016}.

Blood flow biosignals generally describe the characteristics of certain physiological signals according to their spectrum properties in five main frequency bands: endothelial/metabolic (3-20mHz), neurogenic (20-50mHz), myogenic (50-150mHz), respiratory (0.15-0.4Hz) or cardiac (0.4-2.0Hz) \citep{WaveletOsc-Stefanovska-1999, Wavelet-Geyer-2004, PhysicsHumanCardiovascular-Stefanovska-1999}. In this paper, we denoted this frequency classification as the ENMRC (endothelial-neurogenic-myogenic-respiratory-cardiac) scale. The frequency association of this spectral division with vasomotion and neural activity was first investigated by Kastrup et al. \citep{VasomotionHumanSkin-Kastrup-1989} and later verified by Soderstrom et al. \citep{InvolvementSympatheticNerve-Soderstrom-2003}. ENMRC has also been successfully extrapolated to fNIRS for monitoring patients in an intensive care unit \citep{DynamicTrackingMicrovascular-Mendelson-2020} in order to recognize variations in connectivity and muscle exhaustion \citep{DifferencesNetInformation-Urquhart-2020}, cognitive activities \citep{SimulFNIRSThermal-Pinti-2015}, physical activity \citep{EffectCerebralVasomotion-Bosch-2017} or sleep \citep{PredominantEndothelialVasomotor-Zhang-2014}.

The ability to distinguish spectral-specific responses is a primary advantage of the ENMRC frequency division. For instance, the endothelial band denotes the highest wavelet coherence during arithmetic-based tasks \citep{SimulFNIRSThermal-Pinti-2015}. Besides, exercise performance efficiency positively correlates with spectrum power in the neurogenic band while negatively correlated in the endothelial band \citep{PredominantEndothelialVasomotor-Zhang-2014}.

The metabolic, neurogenic and myogenic intervals of the ENMRC frequency division fall within the category of very-low-frequencies \citep{VasomotionHumanSkin-Kastrup-1989}. Estimating the power spectrum density at these frequency intervals is a challenge given that a long sample period is required (an endothelial wave could need up to 333.3 seconds to complete a single cycle). Besides, each spectrum estimation approach may lead to different outcomes with contrasting bias and variances \citep{StatisticalDigitalSig-Hayes-1996}. Furthermore, spectrum estimators such as Welch or autoregressive-based methods cannot provide reliable values for narrow frequencies. Due to these conditions, an alternative approach for defining the spectrum power at localized frequencies is to use an array (or bank) of narrow bandpass, anti-notch filters. Each of the filters in this set can be calibrated to define only one desired frequency with a certain tolerance. To the best of our knowledge, only Folgosi-Correa and Nogueira suggested a method with a similar aim \citep{QuantifyingLowFreqFluctuations-Folgosi-2011}. However, the latter limited their analysis to broader frequency bands (higher than 45mHz) with traditional Butterworth frequency filters without further mathematical treatment.

Based on the filter-bank spectral representation of a stationary signal, we propose a new spectrum estimation approach that relies on first-order, complex, infinite impulse response filters: the complex-pole filter representation (COFRE). We developed several properties from COFRE filters that allow us to configure their frequency resolution (and optimizing frequency peak identification) or enhance their time response (for real-time applications). Our method is used to accurately discover discriminatory frequency signatures on fNIRS signals between patients with tinnitus and healthy control groups. The tinnitus spectral signatures are then interpreted within the ENMRC spectral division that serves us as a biological interpretation framework for the detected spectral differences. In the following sections, we introduce the formal definition of the alternative filter bank representation, the COFRE method and the properties of its filters, and the results of its application on the tinnitus dataset.

\section{COFRE: Complex-pole filter representation}

\subsection{Filter-bank spectral representation}

Most biomedical signals have non-stationary characteristics, i.e., their statistical properties can change over time. Khoa et al.~suggested using Wavelets to model non-stationary signals affected by trends, drifts, or event-based changes \citep{RecognizingBrainActivities-Khoa-2008}. However, under normal circumstances, physiological signals can be successfully approximated as local-stationary in short time intervals \citetext{\citealp[p.~390]{HandbookBrainTheory-Arbib-2003}; \citealp{TemporalDerivativeDistribution-Fishburn-2019}}.

Let us assume, therefore, that an fNIRS signal can be modeled by a wide-sense stationary process (WSS) in a sufficiently small time interval $t\in\left[t_{0},t_{1}\right]$ with local stationarity properties \citep{TemporalDerivativeDistribution-Fishburn-2019}. WSS processes denote a finite and time-invariant first and second statistical moment, with a power spectrum density defined as the Fourier transform of its autocorrelation function $\gamma_{xx}\left(\tau\right)=\mathbb{E}\left[X^{*}\left(t\right)X\left(t+\tau\right)\right]$ \citep[p.~57]{StatisticalDigitalSig-Hayes-1996}:

\begin{equation}%
{
S_{x}\left(\omega\right)
  = \int_{-\infty}^{\infty}
          \gamma_{xx} \left(\tau\right)
          e^{-j2\pi\omega\tau}\,d\tau
}\label{eq:def-psd}%
\end{equation}%

Consequently, given an observed time series $\{x\left(t\right) | t=1,2,\ldots T \}$ of the process $X(t)$, we can define the estimator of the autocorrelation as%
\begin{equation}%
{
\hat{\gamma}_{xx}\left(\tau\right)
  = \frac{1}{T}
     \sum_{t=1}^{T-1-\tau}x^{*}\left(t\right)x\left(t+\tau\right)
  = \frac{1}{T}
     x^{*}(-\tau) \ast x(\tau)
}%
\end{equation}%
where $\ast$ represents the convolution operation, and $x^*$ is the complex conjugate of $x$.

Then, the estimator $\hat{\gamma}_{xx}(\tau)$ can be used to obtain an estimation of the spectrum,%
\begin{equation}%
{
\hat{S}_{x}\left(\omega\right)
  = \sum_{\tau=0}^{T}\hat{\gamma}_{xx}\left(\tau\right)e^{-j2\pi\tau\frac{\omega}{T}}
}\label{eq:spec-def1}%
\end{equation}%
where $\omega$ is the normalized frequency with respect to the sampling frequency $f_{s}$: $\omega=\frac{f}{f_{s}}$, $\omega\in\left[0,\nicefrac{1}{2}\right]$.

Recall that the spectrum can be non-parametrically expressed through the discrete Fourier transform (DFT) of the realization $x(t)$:%
\begin{equation}%
{
\hat{S}_{x}\left(\omega\right)
  = \frac{1}{T} \sum_{\tau=0}^{T}
      x^{*}(-\tau) \ast x(\tau)
      e^{-j2\pi\tau\frac{\omega}{T}}
  = \frac{1}{T}\left|X\left(e^{j2\pi\omega}\right)\right|^{2}
}\label{eq:estimator-spectrum}%
\end{equation}%
where $X\left(e^{j2\pi\omega}\right)$ is the DFT of $x\left(t\right)$. Convergence properties of a tapped-version of \autoref{eq:estimator-spectrum} are discussed in \citep[pp.255-257, Theorem 7.1]{TimeSeriesMixed-Li-2014}.

We should denote that due to the constraints inherited from $X\left(e^{j2\pi\omega}\right)$, $\hat S_X\left(\omega\right)$ has a frequency resolution ($\omega\in\left[0,\nicefrac{1}{2}\right]$):%
\begin{equation}%
{
\Delta_{FFT} \omega = \frac{1}{T}
}\label{eq:FFT-freq-res}%
\end{equation}%

Now, let us define a filtering process $m_\omega(t)$ which spectrum can be described by%
\begin{equation}%
{
M_{w}\left(\omega\right)
= \begin{cases}
    1
  & \omega=\omega^{*}
  \\
  0
  & |\omega-\omega^{*}| \ge \Delta_{FFT}\omega
\end{cases}
}\label{eq:narrow-psd}%
\end{equation}%
with the further restriction that $M_{\omega}(\omega) \le \varepsilon < 1$ where $0 < |\omega-\omega^{*}| \le \Delta_{FFT}\omega$.

Therefore, DFT can be expressed as a sum of a family of narrow-band pass filters $\{ M_{ k \Delta_{FFT}\omega } (\omega)\; \vert k=0,1,\ldots, \lfloor \frac T 2 \rfloor \}$:%
\begin{equation}%
{
\begin{aligned}
X^{f}\left(e^{j2\pi\omega}\right)%
  & = \sum_{k=0}^{\lfloor \frac T 2 \rfloor}%
      M_{k \Delta_{FFT}\omega}(\omega) X\left(e^{j2\pi\omega}\right) 
  & \omega \in W=\left\{0,\Delta_{FFT}\omega,\ldots, \left\lfloor \frac T 2 \right\rfloor \Delta_{FFT}\omega \right\}
\end{aligned}
}\label{eq:spectrum-as-filters}%
\end{equation}%

Note that $M_{\omega}(\omega)$ can also be interpreted as a spectrum smoothing function in the interval $\omega^{*} - \Delta_{FFT}\omega < \omega < \omega^{*} + \Delta_{FFT}\omega$.

Finally, the spectrum in the entire support $W$ can also be expressed as a filter sum:%
\begin{equation}%
{
\hat{S}_{x}\left(\omega\right)
  = \frac{1}{T}\left|X^{f}\left(e^{j2\pi\omega}\right)\right|^{2}
  \quad
  \omega \in W
}%
\end{equation}%

Alternatively, the spectrum at a specific frequency $\omega_0$ can be estimated using a filter $M_{\omega_0}$:%
\begin{equation}%
{
\hat{S}_{x}\left(\omega=\omega_0\right)
  = \frac{1}{T}
  \left|M_{\omega_0}(\omega) \right|^{2}
  \left|X\left(e^{j2\pi\omega_0}\right)\right|^{2}
}\label{eq:filter-spectrum}%
\end{equation}%

\subsection{Complex-pole narrow bandpass filter}

\begin{figure}
\hypertarget{fig:filter-properties}{%
\centering
\includegraphics[width=0.6\textwidth,height=\textheight]{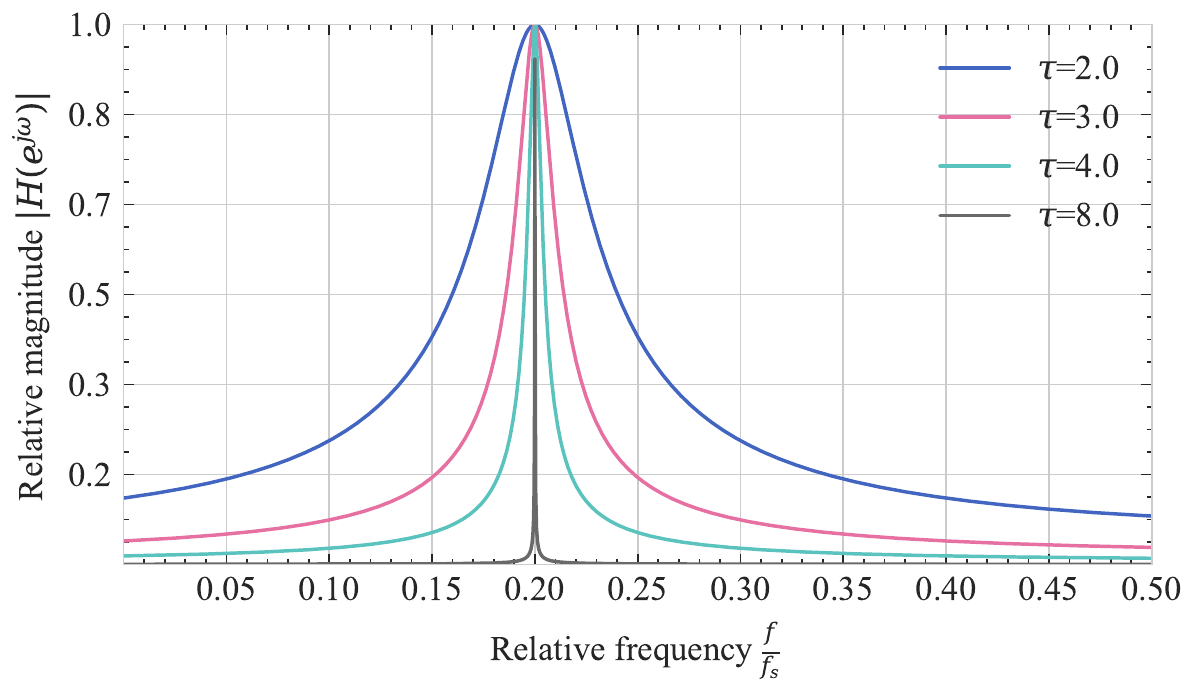}
\caption{Magnitude response $|H \left(e^{j\omega}\right)|$ of a complex narrow-band pass filter when $\omega^*=0.2$.}\label{fig:filter-properties}
}
\end{figure}

The COFRE spectral representation relies in the use of an array of narrow-bandpass filters. These filters $\mathcal H_k$ are infinite impulse response (IIR) filters with a single complex pole by construction.

Given a real-valued input signal $x\left(t\right)$, let us define a complex-output filter $\mathcal{H}$ that maps $x\mapsto y$:%
\begin{equation}%
{
y\left(t\right)=\phi y\left(t-1\right)+x\left(t\right)
}\label{eq:car}%
\end{equation}%
where $y\left(t\right)$ is the complex-valued filtered signal, and $\phi$ is the complex autoregressive coefficient: $\phi=\rho e^{j\omega^{*}}$.

Using the a lag operation $L$ defined as $L^{\ell}x\left(t\right)=x\left(t-\ell\right)$, we can formulate $\mathcal{H}$ as:%
\begin{equation}%
{
\mathcal{H}x\left(t\right) = 
   \left(1-\phi L^{1}\right)^{-1}x\left(t\right)
}\label{eq:filter-def-L}%
\end{equation}%

The intrinsic frequency properties of $\mathcal H$ can be described through its transfer function (TF) in the z-domain. TF can be naturally extracted by inverting \autoref{eq:filter-def-L} after replacing $L=z^{-1}$:

\begin{equation}%
{
H\left(z\right)
  = \frac{1}
         {1-\rho e^{j2\pi\omega^{*}}z^{-1}}
  = F\left(z;\rho,\omega^{*}\right)
}\label{eq:filter-def-z}%
\end{equation}%

In consequence, the filter magnitude response, or gain function, $M(\cdot)$ depends on the system frequency and it is described by%
\begin{equation}%
{
M\left(\omega;\rho,\omega^{*}\right)
  = \left|%
          H \left(e^{j2\pi\omega}\right)%
    \right|
  = \frac{1}
         {\left|1-\rho e^{j2\pi\left(\omega^{*}-\omega\right)}\right|}
}\label{eq:magnitude}%
\end{equation}%
where the $\omega$ is the normalized signal frequency: $\omega=\frac{f}{f_{s}}$, $f$ is the signal frequency in Hz, and $f_{s}$ is the sampling rate of the signal.

The complex autoregressive coefficient $\phi$ reflects some significant characteristics of the filter magnitude response: the filter has a \textbf{symmetric} response around $\omega^*$ where the filter also denotes its maximum gain (\autoref{fig:filter-properties}.A). These properties are formalized in \autoref{lem:conditions} and \autoref{lem:uniqueness}.

\begin{lemma}[Symmetry response]\label{lem:conditions}A complex single-pole IIR filter defined by the transfer function $M\left(\omega;\rho,\omega^{*}\right)$ is symmetric around $\omega^{*}$, i.e, $M\left(\omega;\rho,\omega^{*}-\Delta\omega\right)=M\left(\omega;\rho,\omega^{*}+\Delta\omega\right)\,\Delta \omega\in[0,\frac 1 2]$.\end{lemma}

\begin{proof}Proof in Section~\ref{sec:symmetric-response}.\end{proof}

\begin{lemma}[Unique maximum]\label{lem:uniqueness}The filter $M\left(\omega;\rho,\Delta\omega\right)$ has a single and unique maximum located at $\omega^{*}$ and%
\begin{equation}%
{
\max M\left(\omega;\rho,\omega\right)=M\left(\omega;\rho,\omega^{*}\right)=\left|1-\rho\right|^{-1}
}%
\end{equation}%
\end{lemma}

\begin{proof}Proof in Section~\ref{sec:unique-maximum}.\end{proof}

Let us now reparametrize the coefficient modulus $\rho$ (\autoref{eq:magnitude}) by introducing a new variable: the \emph{frequency bandwidth} $\tau$ such that $\rho=1-e^{-\tau}$. Filters with lower values of $\tau$ have a wide symmetric response around $\omega^*$. In comparison, higher $\tau$ values will induce very narrow filters (\autoref{fig:filter-properties}.A). To ensure the filter is stationary (and therefore, stable), the condition $\left|\phi\right|<1$ should be satisfied, or equivalently, $\left|\rho\right|<1$ or $\tau>0$.

\begin{figure}
\hypertarget{fig:frequency-resolution}{%
\centering
\includegraphics[width=0.8\textwidth,height=\textheight]{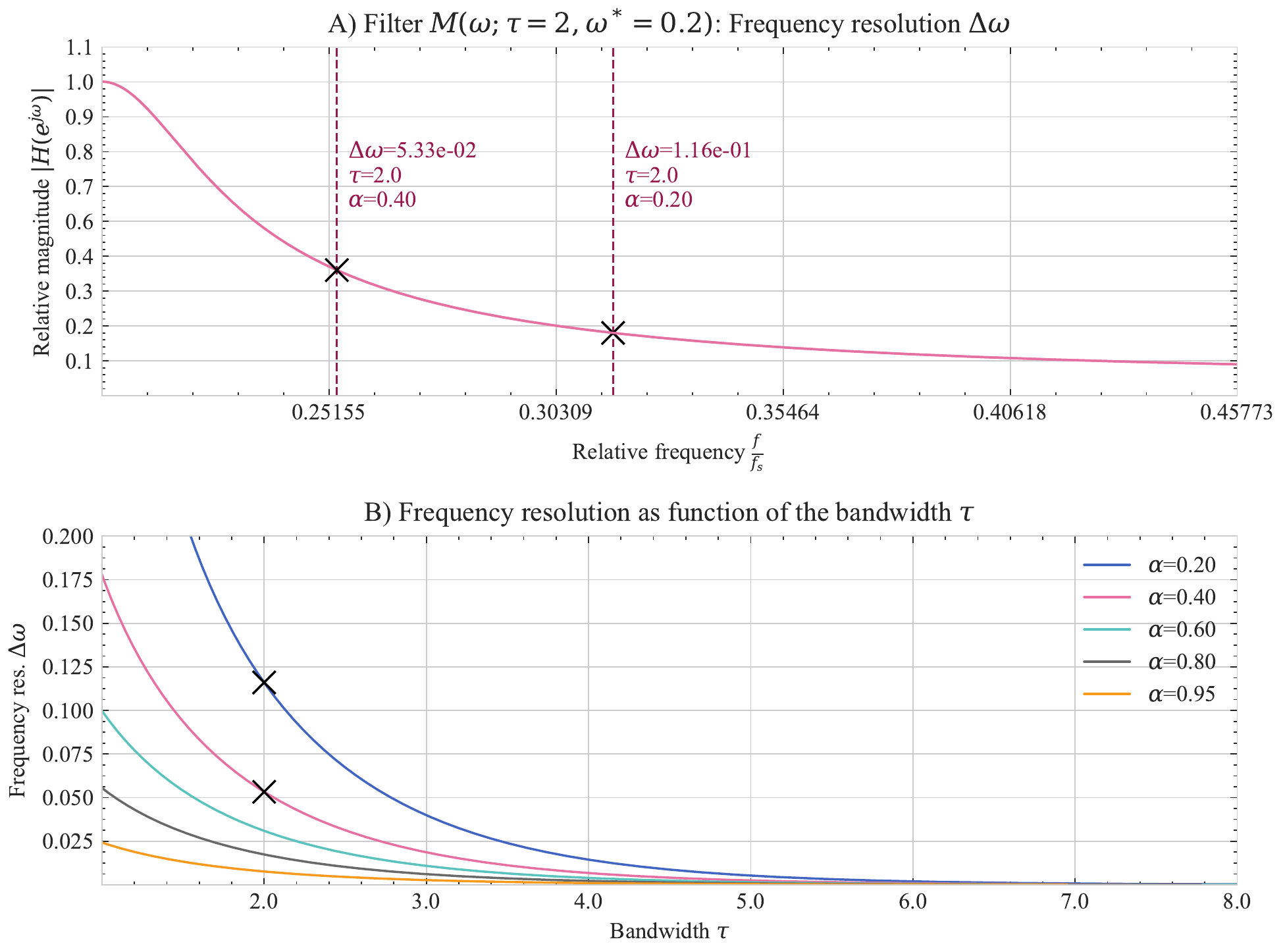}
\caption{Frequency resolution of a complex narrow-band pass filter: A) Visual interpretation of two frequency resolutions with respect to the filter magnitude response (\autoref{def:frequency-resolution}). B) Curves of the frequency resolution with respect to the bandwidth $\tau$ and cut-offs $\alpha$ (\autoref{lem:frequency-resolution}).}\label{fig:frequency-resolution}
}
\end{figure}

\begin{figure}
\hypertarget{fig:rise-time}{%
\centering
\includegraphics[width=0.8\textwidth,height=\textheight]{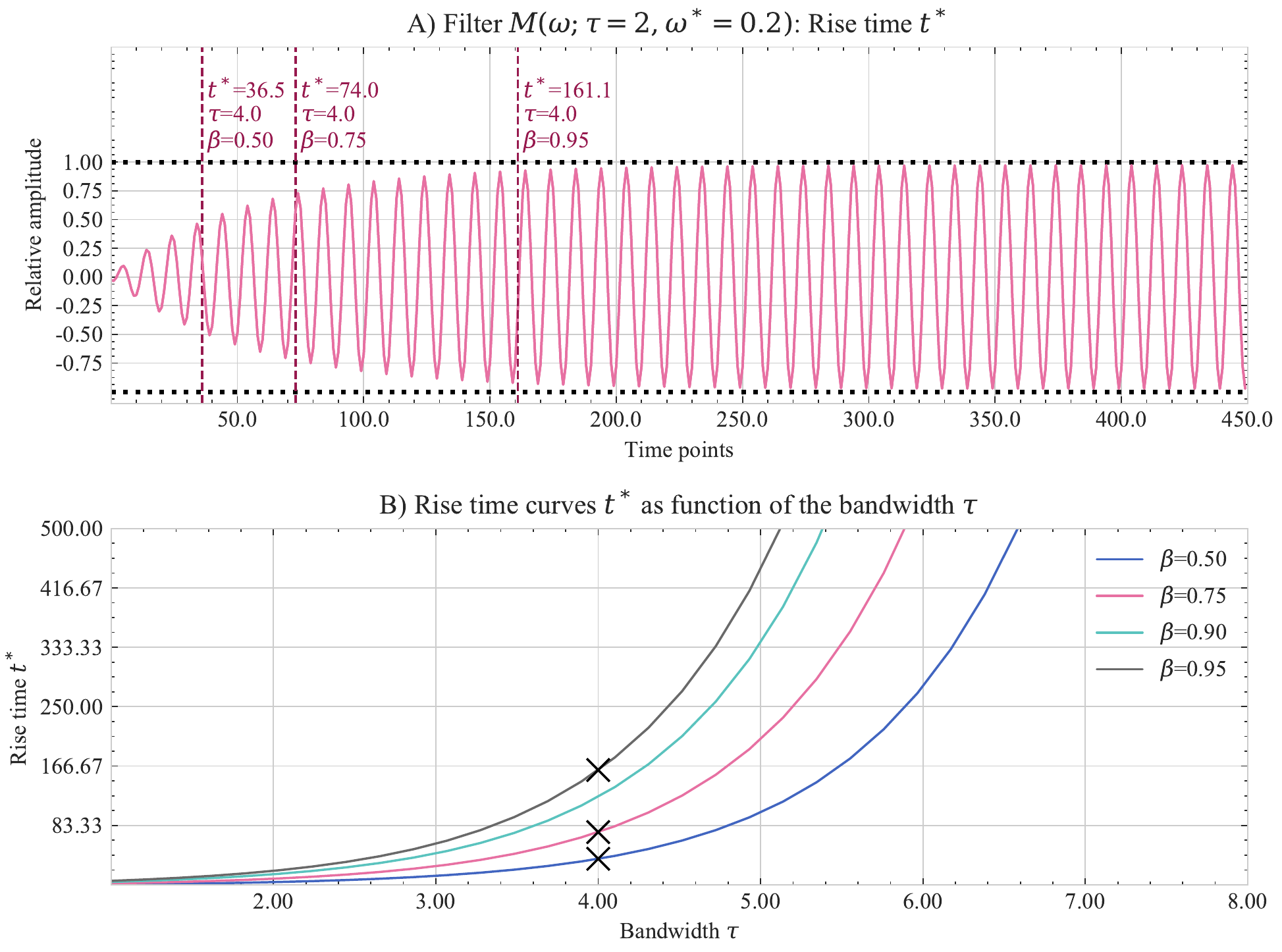}
\caption{Rise-time (transient time response) of a complex narrow-band pass filter: A) Visual interpretation of three rise-time parameters with respect to the transient response of $\mathcal{R}e \bigl\{y\left(t\right)\bigr\}$ with a sinusoid input signal (\autoref{def:rise-time}). B) Curves of the rise time with respect to the bandwidth $\tau$ and cut-offs $\beta$ (\autoref{lem:rise-time}).}\label{fig:rise-time}
}
\end{figure}

\begin{figure}
\hypertarget{fig:time-freq-uncertainty}{%
\centering
\includegraphics[width=0.9\textwidth,height=\textheight]{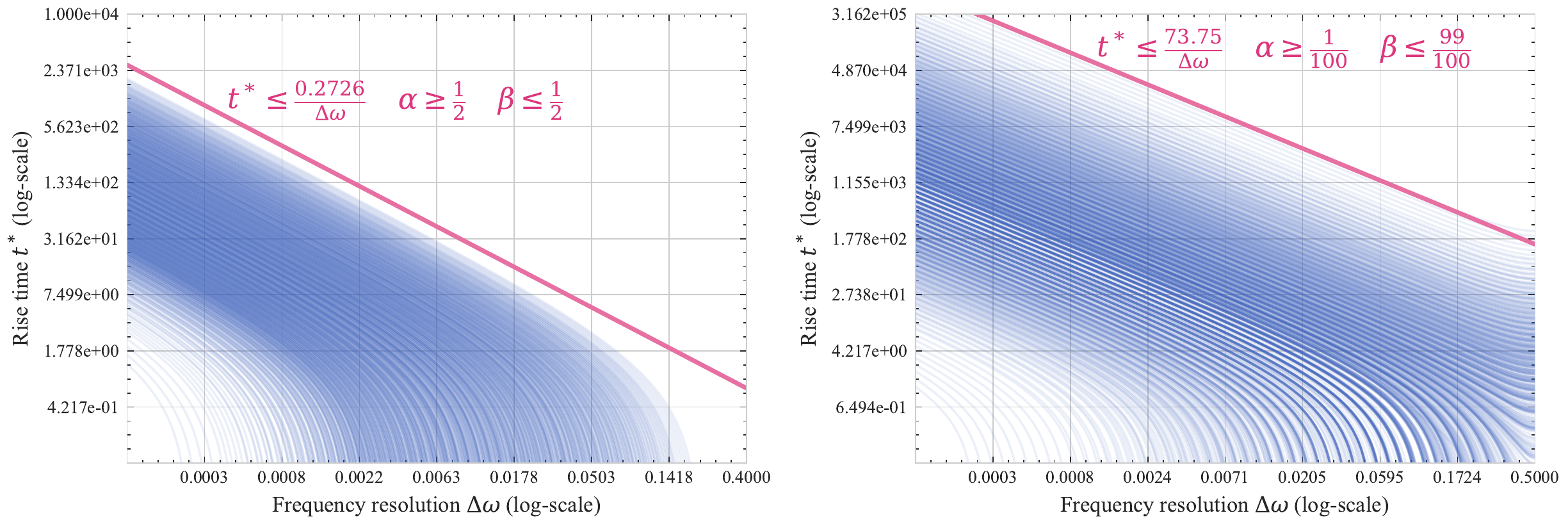}
\caption{Joint time-frequency $(t^*,\Delta\omega)$ curves for several cut-off intervals of $\alpha$ and $\beta$ along with their respective upper-bounds described in \autoref{lem:time-frequency-uncertainty}.}\label{fig:time-freq-uncertainty}
}
\end{figure}

Furthermore, as a direct consequence of \autoref{eq:car} and \autoref{eq:magnitude}, it is remarkable that the filter $F\left(z;\rho,\omega^{*}\right)$ will be conditioned by two main constraints: its maximum frequency resolution and its transient response. For the purpose of this paper, we formalized these concepts with the following definitions:

\begin{definition}[Frequency resolution]\label{def:frequency-resolution}Given a cut-off factor $\alpha\in\left(0,1\right)$, define the frequency resolution $\Delta\omega>0$ as the frequency distance that satisfies%
\begin{equation}%
{
M\left(\omega;\rho,\omega\right)
  \ge \alpha M\left(\omega;\rho,\omega^{*}\right)
  \quad\forall
  \omega\in\left[\max\left(0,\omega^{*}-\Delta\omega\right),\min\left(\pi,\omega^{*}+\Delta\omega\right)\right]
}%
\end{equation}%
\end{definition}

\begin{corollary}[]\label{def:alternative-frequency-resolution}Due to the convexity of $M\left(\omega;\rho,\omega\right)$, the frequency resolution $\Delta\omega$ also satisfies%
\begin{equation}%
{
M\left(\omega;\rho,\omega^*\pm\Delta\omega\right)
  = \alpha M\left(\omega;\rho,\omega^{*}\right)
}%
\end{equation}%
\end{corollary}

\begin{definition}[Rise or transient time]\label{def:rise-time}Given a cut-off $\beta\in\left(0,1\right)$ and an input signal $x\left(t\right)=\cos\left(\omega^{*}t\right)$, define the rise time $t^{*}$ as the time point such that%
\begin{equation}%
{
\mathcal{R}e \bigl\{y\left(t\right)\bigr\} \ge\beta x\left(t\right)
\,\forall t\ge t^{*}
}%
\end{equation}%
\end{definition}

The interpretation of $t^*$ and $\Delta\omega$ is straightforward: the frequency resolution $\Delta\omega$ sets the frequency radius where the filter has a proportion $\alpha$ of its peak response (\autoref{fig:frequency-resolution}). On the time-domain, the rise time $t^*$ estimates the number of points that the filter requires to start providing a steady-state output (\autoref{fig:rise-time}). Both constraints have closed-form expressions that depend only on the filter bandwidth $\tau$ and central frequency $\omega^*$ as the following lemmas and corollaries denote.

\begin{lemma}[Frequency resolution configuration]\label{lem:frequency-resolution}The cosine of the frequency resolution of a filter $M\left(\omega;\rho,\omega^{*}\right)$ is a second-order rational function of the complex autoregressive modulus $\rho$:%
\begin{equation}%
{
\begin{aligned}
\Delta\omega
&= \frac{1}{2\pi}\arccos
   \left(\frac{1}{2\rho}
  \left(\rho^{2}+1-\alpha^{-2}\left(1-\rho\right)^{2}\right)
   \right)
\end{aligned}
}%
\end{equation}%
\end{lemma}

\begin{proof}Proof in Section~\ref{sec:frequency-resolution}.\end{proof}

\begin{corollary}[Frequency-optimal bandwidth]\label{lem:frequency-optimal-bandwidth}The minimum bandwidth $\tau$ to ensure a frequency resolution $\Delta\omega$ under a cut-off $\alpha$ is given by\end{corollary}

\begin{equation}%
{
\begin{aligned}
\tau
  &= -\log\left(
        1
        - \frac{\cos2\pi\Delta\omega-\alpha^{-2}}{1-\alpha^{-2}}
        - \frac{1}{1-\alpha^{-2}}
           \sqrt{\cos^{2}2\pi\Delta\omega-\left(1-\alpha^{-2}\right)^{2}}
      \right)
\end{aligned}
}%
\end{equation}%

\begin{proof}Proof in Section~\ref{sec:frequency-optimal-bandwidth}.\end{proof}

\begin{lemma}[Rise time configuration]\label{lem:rise-time}The rise time of a filter $M\left(\omega;\rho,\omega^{*}\right)$ is inversely proportional to the logarithm of $\rho$ and proportional to the logarithm of the complement of the cut-off factor $\beta$:%
\begin{equation}%
{
t^{*}=\log_{\rho}\left(\frac{1-\beta}{\rho}\right)=\frac{\log\left(1-\beta\right)}{\log\left(\rho\right)}-1
}%
\end{equation}%
\end{lemma}

\begin{proof}Proof in Section~\ref{sec:rise-time}.\end{proof}

\begin{corollary}[Time-optimal bandwidth]\label{lem:rise-optimal-bandwidth}The minimum bandwidth $\tau$ to ensure a rise time $t^{*}$ (under a cut-off $\beta$) is%
\begin{equation}%
{
\tau=-\log\left(1-\left(1-\beta\right)^{\frac{1}{1+t^{*}}}\right)
}%
\end{equation}%
\end{corollary}

\begin{proof}Proof in Section~\ref{sec:rise-optimal-bandwidth}.\end{proof}

\begin{lemma}[Joint time-frequency constraint]\label{lem:time-frequency-uncertainty}Given the cut-offs $\alpha$ and $\beta$, the rise time $t^{*}$ and the frequency resolution $\Delta\omega$ are mutually constrained by%
\begin{equation}%
{
t^{*}\Delta\omega
   < -\log\left(
    1-\beta\right)
 \left(\frac{\sqrt{\left(\alpha\pi\Delta\omega\right)^{2}-\alpha^{2}+1}}{2\pi\alpha}+\frac{\Delta\omega}{2}\right)
}%
\end{equation}%
\end{lemma}

\begin{proof}Proof in Section~\ref{sec:time-frequency-uncertainty}.\end{proof}

\begin{corollary}[Time-frequency uncertainty]\label{lem:time-frequency-uncertainty-summary}A narrow bandpass filter with a frequency resolution $\Delta\omega<\frac 1 5$ has its joint time-frequency resolution upper-bounded by%
\begin{equation}%
{
t^{*}\Delta\omega<-\log\left(1-\beta\right)\left(\frac{\sqrt{1-\left(1-\frac{4}{25}\pi^{2}\right)\alpha^{2}}}{2\pi\alpha}+\frac{1}{10}\right)
}%
\end{equation}%
\end{corollary}

\begin{proof}Proof in Section~\ref{sec:time-frequency-uncertainty}.\end{proof}

A graphical representation of the boundaries described in \autoref{lem:time-frequency-uncertainty} is shown in \autoref{fig:time-freq-uncertainty}.

\subsection{COFRE spectrum representation}

From \autoref{eq:spectrum-as-filters}, it is established that the spectrum of a real-valued input signal $x\left(t\right)$ can be expressed as the sum of narrow bandpass filters with frequency resolution $D=\Delta_{FFT}\omega$ (\autoref{eq:FFT-freq-res}). We use this property to define an array of single-pole filters $\left\{ F_{k}(\omega),\, k=1,2,\cdots,\left\lfloor \frac T 2 \right\rfloor \right\}$ such that%
\begin{equation}%
{
F_{k} \left(\omega\right)
  = F\left(\omega;\rho=1-e^{-\tau},\omega^{*}=kD\right)
}%
\end{equation}%
where the bandwidth $\tau$ is adjusted to the aimed frequency resolution (\autoref{lem:frequency-optimal-bandwidth}).

Furthermore, it is acknowledged (\autoref{eq:filter-spectrum}) that the spectrum at a particular frequency $\omega_0=k_0 D$ can be estimated through a unitary gain filter with a central frequency in $\omega_0$:%
\begin{equation}%
{
\hat{S}_{x}\left(\omega=k_0 D\right)
  = \frac{1}{T}
  \left|\frac{1}{1-\rho} F_{k_0}(\omega)
  X\left(e^{j2\pi k_0 D}\right)\right|^{2}
}\label{eq:filter-spectrum-f}%
\end{equation}%
where $\frac{1}{1-\rho} F_{k_0}(\omega)$ has a unit gain at $\omega_0$: $\left\vert\frac{1}{1-\rho} F_{k_0}(\omega_0)\right\vert = \frac{1}{1-\rho} \left\vert 1 - \rho \right\vert=1$.

Note that the inverse Fourier transform of $Y(\omega) = F_{k_0}(\omega) X\left(e^{j2\pi k_0 D}\right)$ is given by \autoref{eq:car}:%
\begin{equation}%
{
y_{k_0}\left(t\right)
  = \mathcal{H}^{kD} x\left(t\right)
  = \rho e^{-jk_0 D} y_{k_0} \left(t-1\right)+x\left(t\right)
}%
\end{equation}%

Recall the Parseval's identity, $\sum_{t=1}^{T}\left|x\left(t\right)\right|^{2}=\int_{-\pi}^{\pi}\left|X\left(e^{j2\pi\omega}\right)\right|^{2}d\omega$, then%
\begin{equation}%
{
\sum_{t=1}^{T}\left|y\left(t\right)\right|^{2}
   = \int_{-\pi}^{\pi}\left|Y\left(e^{j2\pi\omega}\right)\right|^{2}d\omega
   \approx \left|Y\left(e^{j2\pi kD}\right)\right|^{2}
}%
\end{equation}%

Alternatively, \autoref{eq:filter-spectrum-f} can be rewritten as%
\begin{equation}%
{
\hat{S}_{x}\left(\omega=k_0 D\right)
  = \frac{1}{\left(1-\rho\right)^2}
  \, \frac{1}{T}
  \left|Y\left(e^{j2\pi kD}\right)\right|^{2}
  \approx
  \frac{1}{\left(1-\rho\right)^2}
  \, \frac{1}{T}
  \sum_{t=1}^{T}\left|y\left(t\right)\right|^{2}
}\label{eq:spectrum-var}%
\end{equation}%

Then, we can formulate a spectrum estimator at a frequency $\omega^{*}=kD$ using the biased variance estimator of the filtered signal at the same frequency, for a sufficiently narrow bandwidth $\tau=-\log\left(1-\rho\right)$:

\begin{equation}%
{
\hat{S}_{x}\left(\omega^{*}\right)
=
\frac{1}{e^{-2\tau}}
  \text{var}\left(\mathcal{H}^{kD}x\left(t\right)\right)
}\label{eq:spectrum-estimator}%
\end{equation}%

\subsection{Recursive COFRE estimation}

Let us remark $\hat{S}^{(T)}_{x}\left(\omega=k_0 D\right)$ as the spectrum estimator obtain from $T$ points (\autoref{eq:spectrum-var}):%
\begin{equation}%
{
\hat{S}^{(T)}_{x}\left(\omega=k_0 D\right)
  =
  \frac{1}{e^{-2\tau}}
  \, \frac{1}{T}
  \sum_{t=1}^{T}\left|y\left(t\right)\right|^{2}
}%
\end{equation}%

Therefore, it is straightforward to estimate the spectrum using $t-1$ points recursively:%
\begin{equation}%
{
\hat{S}^{(t)}_{x}\left(\omega=k_0 D\right)
  =
  \frac{t-1}{t}
  \hat{S}^{(t-1)}_{x}\left(\omega\right)
  +
  \frac{1}{t\, e^{-2\tau}}
  \left| \mathcal{H}^{k_0D}x\left(t\right) \right|^{2}
}%
\end{equation}%

Remark that through the autoregressive formulation in \autoref{eq:car}, $y(t)$ only depends on the previous filtered point $y(t-1)$ and the current new point $x(t)$:%
\begin{equation}%
{
\mathcal{H}^{k_0D}x\left(t\right)
= \rho e^{-jk_0 D} \mathcal{H}^{k_0D}x\left(t-1\right)+x\left(t\right)
}%
\end{equation}%

We should emphasize that this recursive formulation allows a proper estimation of the spectrum for any observation $x(t)$ collected after the rise time $t>t^*$ (\autoref{def:rise-time}).

Furthermore, it is evident that this recursive estimation has space and time complexity $\mathcal{O}\left(1\right)$.

\section{Real fNIRS data}

In this paper, we use the data provided in \citep{TinnitusAltersResting-SanJuan-2017}. This dataset was collected at the University of Michigan to analyze brain connectivity changes through fNIRS signals, between patients with tinnitus (PT) and a healthy control (HC) group. The study comprised 20Hz-sampled signals in the frontotemporal cortex during the exposition of PT and HC groups to auditory stimuli. Participants included eight recruited healthy subjects (62\% males) with an average age of 25.4 years, as well as ten adults with subjective bilateral tinnitus with an average age of 48.7 years (60\% males). HC and PT subjects performed three auditory tests to ensure similar physiological conditions: speech reception threshold tests, audiogram exams, and word recognition score tests, that shows no significant difference between the groups discarding a possible objective tinnitus in the PT group.

The experimental procedure consisted of introducing the participants to alternating blocks of silence and sound lasting 18 seconds each period. Auditory stimuli (27 blocks) were evenly divided (and randomly distributed) between three types of stimulus: (a.) single-frequency 700Hz-wave, (b.) single-frequency 8KHz-wave, and (c.) broadband noise. The dataset was recorded using two light sources with wavelengths at 690 nm and 830 nm at a sampling frequency of 20 Hz. For further description of the experimental protocol, we refer to \citep{TinnitusAltersResting-SanJuan-2017}.

We use this dataset to identify variations in the overall spectrum between PT and HC groups. For our analysis, we focus on the electrodes T3 and T4 located in the left and right side of the temporal lobe (auditory cortex). Optical intensities from these locations were converted into chromophore concentration changes using the modified Beer-Lambert law:

\begin{equation}%
{
\left(\begin{array}{c}
x_{HbR}\\
x_{HbO}
\end{array}\right)=\frac{1}{L}\left(\left(\begin{array}{cc}
\text{DPF}_{\lambda_{690}} & 0\\
0 & \text{DPF}_{\lambda_{830}}
\end{array}\right)\left(\begin{array}{cc}
\varepsilon_{690,HbR} & \varepsilon_{690,HbO}\\
\varepsilon_{830,HbR} & \varepsilon_{830,HbO}
\end{array}\right)\right)^{-1}\left(\begin{array}{c}
\Delta\mu_{\lambda_{690}}\\
\Delta\mu_{\lambda_{830}}
\end{array}\right)
}%
\end{equation}%

where $\Delta\mu_\lambda$ are the changes in absorption in a wavelength $\lambda$ \citep[p.~8]{RvwInfrared-Jue-2013} , $x_{c}$ are the estimates of the concentration changes of a chromophore $c$, $\varepsilon_{\lambda, c}$ are the molar extinction coefficients, $\text{DPF}_{\lambda}$ are the differential pathlength factor for the two wavelengths, and $\circ$ is the Hadamard product operator. For this study, we use $\varepsilon_{690,HbR}=2051.96$, $\varepsilon_{830,HbR}=693.4$, $\varepsilon_{690,HbO}=276.0$, $\varepsilon_{830,HbO}=974.0$ as defined in \citep{OpticalBlood-Schmitt-1986, MultipleScatteringField-Moaveni-1970}. Moreover, we set $\text{DPF}_{\lambda_{690}}=6.51$ and $\text{DPF}_{\lambda_{830}}=5.86$ as recommended by \citep{CranialOptical-Duncan-1996} for groups with a similar age than the participants in our dataset.

For conciseness in the analysis, let us define the sound spectral contrast $c(\omega)$. This metric is defined as the difference (for a group $g=\{HC,PT\}$) of the estimated spectrum during a sound stimulus and the spectrum estimated during the followed silence period:%
\begin{equation}%
{
c\left(\omega\vert g\right)
  = \bigl\vert
    {S}_{x}\bigl(\omega\vert\text{sound},g\bigr)
  - {S}_{x}\bigl(\omega\vert\text{silence},g\bigr)
  \bigr\vert
}\label{eq:sound-silence-contrast}%
\end{equation}%

We assume that ${S}_{x}\left(\omega\vert\text{sound},g\right)$ and ${S}_{x}\left(\omega\vert\text{silence},g\right)$ are uncorrelated. Therefore, $c^2(\omega_i\vert g)\perp c^2(\omega_j\vert g)\;\forall i\ne j$,

Now, let us define the inter-group contrast $d(\omega)$ as the difference between the sound spectral contrast in the TP and HC:%
\begin{equation}%
{
d\left(\omega\right)
  = c\left(\omega\vert\text{TP}\right)
  - c\left(\omega\vert\text{HC}\right)
}\label{eq:group-contrast}%
\end{equation}%

Due to the uncorrelatedness property of $c(\omega\vert g)$, and the central limit theorem, it is known that the limiting distribution of the estimator for the expected value $\mathbb{E}[d(\omega)]=\bar d(\omega)$ follows a normal distribution:%
\begin{equation}%
{
\bar d\left(\omega\right)
  \sim\mathcal{N}\left(
    d_{0}\left(\omega\right),
    \sigma_{d}^{2}\left(\omega\right)
  \right)
}\label{eq:limiting-distribution}%
\end{equation}%

\section{Results and discussion}

\begin{figure}
\hypertarget{fig:spectrum-contrast}{%
\centering
\includegraphics[width=1\textwidth,height=\textheight]{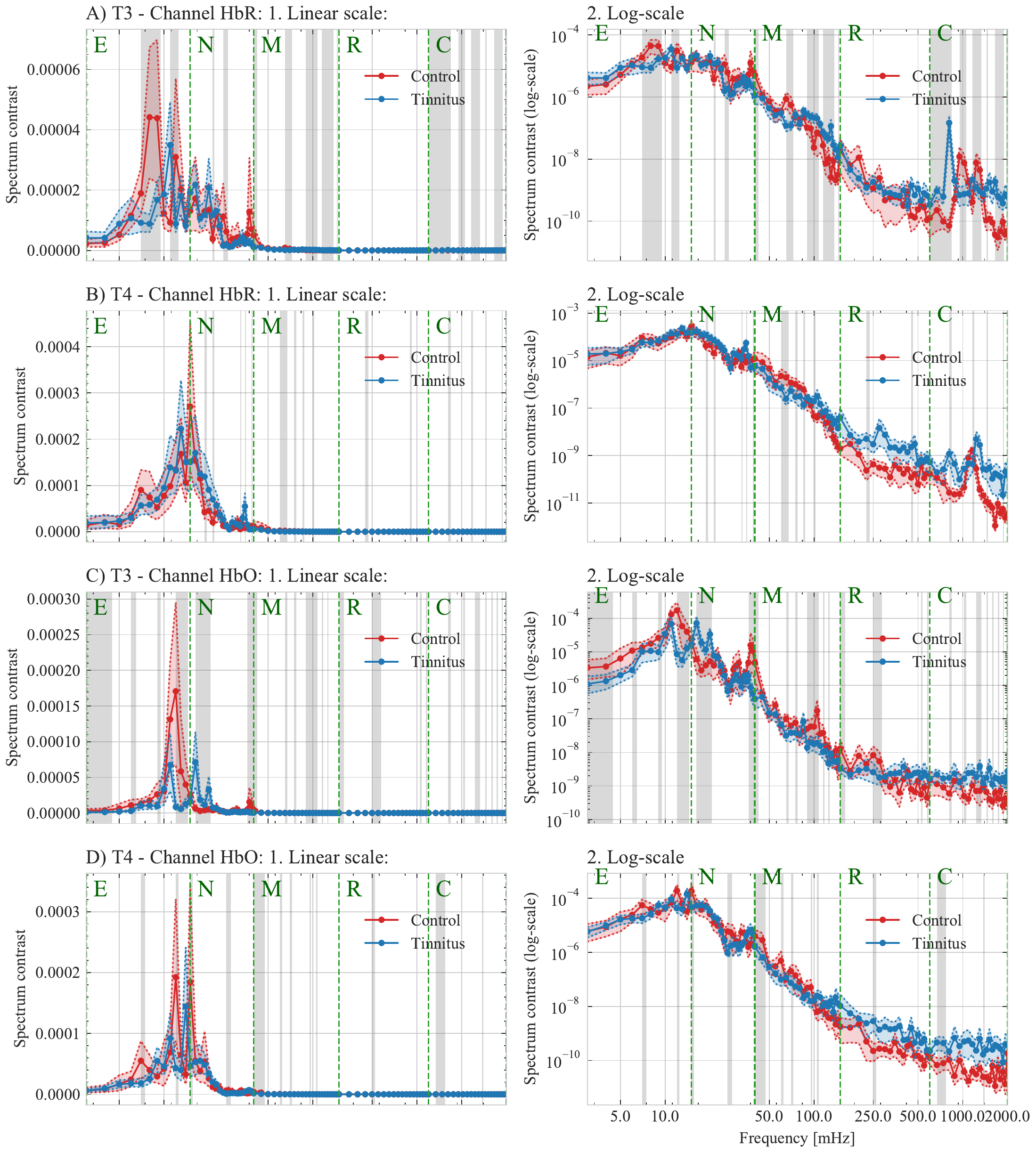}
\caption{Sound spectrum contrast $c(\omega)$ for patients with tinnitus (PT) and the healthy control (HC) group in channels T3 and T4 for oxy- and deoxy-hemoglobin. Mean values are shown in bold lines, while dotted lines denote the 5th and 95th percentile of the estimates' distribution. Intervals of the ENMRC (endothelial, neurogenic, myogenic, respiratory and cardiac) scale are shown in dotted green lines. Black shadow regions denote the spectrum intervals with significant spectrum difference $d(\omega)$ between the PT and HC ($\text{p-value}<0.05$).}\label{fig:spectrum-contrast}
}
\end{figure}

\begin{table}\hypertarget{tab:significant-regions}{%
\centering
\scalebox{0.6}{\begin{minipage}[t]{19cm}%
\begin{tabular}{lllllll}
\toprule
\textbf{Hb} & \textbf{Ch.} & \textbf{ENMRC} & \textbf{Frequency} & \textbf{Inter-group contrast} & \multicolumn{2}{l}{\textbf{Significance (maxP method)}}\tabularnewline
 &  & \textbf{scale} & \textbf{interval} & Mean, 95\% C.I. & Min. t-value & Max. p-value\tabularnewline
\hline
 HbR  &  T3  &  E  &  {[}7.000, 9.000)  &  1.599e-03; {[}1.073e-03, 2.106e-03{]}  &  3.032  &  2.694e-03 {*}{*} \tabularnewline
   &    &  E  &  {[}11.000, 13.000)  &  -1.361e-05; {[}-1.436e-03, 1.445e-03{]}  &  -2.688  &  7.684e-03 {*}{*} \tabularnewline
   &    &  N  &  {[}21.250, 22.500)  &  -7.732e-04; {[}-7.732e-04, -7.732e-04{]}  &  -2.422  &  1.619e-02 {*} \tabularnewline
   &    &  N  &  {[}25.000, 27.500)  &  6.203e-04; {[}4.564e-04, 7.831e-04{]}  &  2.395  &  1.736e-02 {*} \tabularnewline
   &    &  N  &  {[}35.000, 36.250)  &  3.759e-04; {[}3.759e-04, 3.759e-04{]}  &  2.380  &  1.811e-02 {*} \tabularnewline
   &    &  NM  &  {[}37.500, 45.000)  &  6.343e-04; {[}4.299e-04, 8.744e-04{]}  &  2.284  &  2.323e-02 {*} \tabularnewline
   &    &  M  &  {[}50.000, 55.000)  &  1.560e-04; {[}1.560e-04, 1.560e-04{]}  &  1.996  &  4.708e-02 {*} \tabularnewline
   &    &  M  &  {[}65.000, 75.000)  &  2.849e-04; {[}2.346e-04, 3.350e-04{]}  &  4.352  &  1.993e-05 {*}{*}{*}{*} \tabularnewline
   &    &  MR  &  {[}90.000, 150.000)  &  -1.147e-04; {[}-1.795e-04, -4.818e-05{]}  &  -2.891  &  4.185e-03 {*}{*} \tabularnewline
   &    &  R  &  {[}250.000, 275.000)  &  -1.347e-05; {[}-1.347e-05, -1.347e-05{]}  &  -3.050  &  2.541e-03 {*}{*} \tabularnewline
   &    &  R  &  {[}300.000, 350.000)  &  -9.696e-06; {[}-1.090e-05, -8.461e-06{]}  &  -2.833  &  4.996e-03 {*}{*} \tabularnewline
   &    &  R  &  {[}375.000, 425.000)  &  -6.766e-06; {[}-8.224e-06, -5.294e-06{]}  &  -2.288  &  2.301e-02 {*} \tabularnewline
   &    &  RC  &  {[}450.000, 1090.000)  &  -1.233e-05; {[}-3.202e-05, 1.101e-06{]}  &  -2.165  &  3.136e-02 {*} \tabularnewline
   &    &  C  &  {[}1160.000, 1230.000)  &  -1.543e-05; {[}-1.543e-05, -1.543e-05{]}  &  -3.600  &  3.864e-04 {*}{*}{*} \tabularnewline
   &    &  C  &  {[}1300.000, 1370.000)  &  1.226e-05; {[}1.226e-05, 1.226e-05{]}  &  2.248  &  2.550e-02 {*} \tabularnewline
 &    &  C  &  {[}1440.000, 2000.000)  &  -1.200e-05; {[}-1.511e-05, -9.196e-06{]}  &  -3.528  &  5.011e-04 {*}{*}{*} \tabularnewline
\cmidrule{2-7} \cmidrule{3-7} \cmidrule{4-7} \cmidrule{5-7} \cmidrule{6-7} \cmidrule{7-7} 
 &  T4  &  E  &  {[}7.000, 8.000)  &  1.949e-03; {[}1.949e-03, 1.949e-03{]}  &  2.283  &  2.333e-02 {*} \tabularnewline
   &    &  EN  &  {[}15.000, 16.250)  &  3.279e-03; {[}3.279e-03, 3.279e-03{]}  &  2.164  &  3.141e-02 {*} \tabularnewline
   &    &  N  &  {[}18.750, 20.000)  &  -2.187e-03; {[}-2.187e-03, -2.187e-03{]}  &  -2.253  &  2.515e-02 {*} \tabularnewline
   &    &  N  &  {[}21.250, 22.500)  &  -2.397e-03; {[}-2.397e-03, -2.397e-03{]}  &  -3.278  &  1.198e-03 {*}{*} \tabularnewline
   &    &  N  &  {[}32.500, 37.500)  &  -1.465e-03; {[}-2.209e-03, -7.654e-04{]}  &  -2.163  &  3.151e-02 {*} \tabularnewline
   &    &  NM  &  {[}38.750, 40.000)  &  7.618e-04; {[}7.618e-04, 7.618e-04{]}  &  2.609  &  9.652e-03 {*}{*} \tabularnewline
   &    &  M  &  {[}60.000, 90.000)  &  2.963e-04; {[}1.662e-04, 4.326e-04{]}  &  2.564  &  1.096e-02 {*} \tabularnewline
   &    &  M  &  {[}100.000, 105.000)  &  -8.845e-05; {[}-8.845e-05, -8.845e-05{]}  &  -2.105  &  3.633e-02 {*} \tabularnewline
   &    &  M  &  {[}130.000, 135.000)  &  -4.033e-05; {[}-4.033e-05, -4.033e-05{]}  &  -2.045  &  4.199e-02 {*} \tabularnewline
   &    &  MR  &  {[}140.000, 350.000)  &  -2.747e-05; {[}-4.112e-05, -1.417e-05{]}  &  -2.106  &  3.624e-02 {*} \tabularnewline
   &    &  R  &  {[}375.000, 525.000)  &  -1.058e-05; {[}-1.302e-05, -7.467e-06{]}  &  -2.320  &  2.119e-02 {*} \tabularnewline
   &    &  R  &  {[}550.000, 575.000)  &  -5.152e-06; {[}-5.152e-06, -5.152e-06{]}  &  -2.149  &  3.263e-02 {*} \tabularnewline
   &    &  C  &  {[}810.000, 1020.000)  &  -6.242e-06; {[}-1.242e-05, -2.233e-06{]}  &  -2.052  &  4.127e-02 {*} \tabularnewline
   &    &  C  &  {[}1090.000, 1230.000)  &  8.663e-06; {[}4.478e-06, 1.289e-05{]}  &  2.176  &  3.052e-02 {*} \tabularnewline
   &    &  C  &  {[}1300.000, 1720.000)  &  -5.317e-06; {[}-9.122e-06, -2.663e-06{]}  &  -2.272  &  2.396e-02 {*} \tabularnewline
   &    &  C  &  {[}1790.000, 2000.000)  &  -2.784e-06; {[}-3.516e-06, -1.679e-06{]}  &  -2.432  &  1.574e-02 {*} \tabularnewline
\midrule
 HbO  &  T3  &  E  &  {[}3.000, 10.000)  &  8.066e-04; {[}1.148e-04, 1.503e-03{]}  &  2.289  &  2.295e-02 {*} \tabularnewline
   &    &  EN  &  {[}11.000, 15.000)  &  2.715e-03; {[}1.097e-03, 4.269e-03{]}  &  2.164  &  3.148e-02 {*} \tabularnewline
   &    &  N  &  {[}16.250, 21.250)  &  -1.551e-03; {[}-2.182e-03, -9.490e-04{]}  &  -2.379  &  1.816e-02 {*} \tabularnewline
   &    &  N  &  {[}26.250, 27.500)  &  2.396e-04; {[}2.396e-04, 2.396e-04{]}  &  2.407  &  1.686e-02 {*} \tabularnewline
   &    &  NM  &  {[}37.500, 45.000)  &  8.031e-04; {[}6.055e-04, 1.012e-03{]}  &  3.056  &  2.496e-03 {*}{*} \tabularnewline
   &    &  M  &  {[}65.000, 70.000)  &  8.213e-05; {[}8.213e-05, 8.213e-05{]}  &  3.419  &  7.371e-04 {*}{*}{*} \tabularnewline
   &    &  M  &  {[}90.000, 100.000)  &  4.988e-05; {[}3.977e-05, 5.995e-05{]}  &  2.545  &  1.154e-02 {*} \tabularnewline
   &    &  M  &  {[}105.000, 110.000)  &  1.130e-04; {[}1.130e-04, 1.130e-04{]}  &  3.472  &  6.128e-04 {*}{*}{*} \tabularnewline
   &    &  M  &  {[}115.000, 120.000)  &  4.652e-05; {[}4.652e-05, 4.652e-05{]}  &  2.984  &  3.138e-03 {*}{*} \tabularnewline
   &    &  MR  &  {[}150.000, 175.000)  &  2.245e-05; {[}2.245e-05, 2.245e-05{]}  &  2.579  &  1.051e-02 {*} \tabularnewline
   &    &  R  &  {[}375.000, 425.000)  &  -1.612e-05; {[}-2.373e-05, -8.605e-06{]}  &  -2.620  &  9.342e-03 {*}{*} \tabularnewline
   &    &  R  &  {[}450.000, 500.000)  &  -1.236e-05; {[}-1.265e-05, -1.207e-05{]}  &  -2.600  &  9.897e-03 {*}{*} \tabularnewline
   &    &  R  &  {[}525.000, 550.000)  &  -1.173e-05; {[}-1.173e-05, -1.173e-05{]}  &  -3.134  &  1.936e-03 {*}{*} \tabularnewline
   &    &  RC  &  {[}575.000, 600.000)  &  -1.519e-05; {[}-1.519e-05, -1.519e-05{]}  &  -3.907  &  1.215e-04 {*}{*}{*} \tabularnewline
   &    &  C  &  {[}670.000, 880.000)  &  -1.186e-05; {[}-1.429e-05, -9.391e-06{]}  &  -2.318  &  2.127e-02 {*} \tabularnewline
   &    &  C  &  {[}1020.000, 1370.000)  &  -1.449e-05; {[}-1.861e-05, -1.084e-05{]}  &  -2.269  &  2.412e-02 {*} \tabularnewline
 &    &  C  &  {[}1440.000, 2000.000)  &  -1.179e-05; {[}-1.393e-05, -9.765e-06{]}  &  -2.189  &  2.953e-02 {*} \tabularnewline
\cmidrule{2-7} \cmidrule{3-7} \cmidrule{4-7} \cmidrule{5-7} \cmidrule{6-7} \cmidrule{7-7} 
 &  T4  &  E  &  {[}7.000, 8.000)  &  1.719e-03; {[}1.719e-03, 1.719e-03{]}  &  2.836  &  4.953e-03 {*}{*} \tabularnewline
   &    &  E  &  {[}12.000, 14.000)  &  2.553e-03; {[}1.100e-03, 4.021e-03{]}  &  2.354  &  1.938e-02 {*} \tabularnewline
   &    &  EN  &  {[}15.000, 16.250)  &  2.757e-03; {[}2.757e-03, 2.757e-03{]}  &  2.490  &  1.346e-02 {*} \tabularnewline
   &    &  N  &  {[}20.000, 21.250)  &  -1.119e-03; {[}-1.119e-03, -1.119e-03{]}  &  -1.996  &  4.708e-02 {*} \tabularnewline
   &    &  N  &  {[}26.250, 28.750)  &  4.849e-04; {[}4.363e-04, 5.318e-04{]}  &  2.162  &  3.163e-02 {*} \tabularnewline
   &    &  N  &  {[}31.250, 33.750)  &  4.020e-04; {[}1.104e-04, 6.965e-04{]}  &  2.165  &  3.134e-02 {*} \tabularnewline
   &    &  NM  &  {[}40.000, 50.000)  &  3.907e-04; {[}2.575e-04, 5.631e-04{]}  &  2.010  &  4.558e-02 {*} \tabularnewline
   &    &  M  &  {[}55.000, 65.000)  &  1.283e-04; {[}1.031e-04, 1.527e-04{]}  &  1.994  &  4.727e-02 {*} \tabularnewline
   &    &  M  &  {[}70.000, 75.000)  &  1.211e-04; {[}1.211e-04, 1.211e-04{]}  &  3.025  &  2.756e-03 {*}{*} \tabularnewline
   &    &  M  &  {[}105.000, 110.000)  &  -3.801e-05; {[}-3.801e-05, -3.801e-05{]}  &  -2.529  &  1.209e-02 {*} \tabularnewline
   &    &  M  &  {[}125.000, 130.000)  &  -2.070e-05; {[}-2.070e-05, -2.070e-05{]}  &  -2.045  &  4.195e-02 {*} \tabularnewline
   &    &  M  &  {[}140.000, 145.000)  &  -3.091e-05; {[}-3.091e-05, -3.091e-05{]}  &  -2.075  &  3.904e-02 {*} \tabularnewline
   &    &  R  &  {[}225.000, 275.000)  &  -1.374e-05; {[}-1.645e-05, -1.103e-05{]}  &  -2.206  &  2.833e-02 {*} \tabularnewline
   &    &  R  &  {[}325.000, 375.000)  &  -8.603e-06; {[}-9.458e-06, -7.747e-06{]}  &  -2.137  &  3.363e-02 {*} \tabularnewline
   &    &  R  &  {[}400.000, 425.000)  &  -5.835e-06; {[}-5.835e-06, -5.835e-06{]}  &  -2.516  &  1.254e-02 {*} \tabularnewline
   &    &  R  &  {[}500.000, 575.000)  &  -6.254e-06; {[}-8.632e-06, -4.621e-06{]}  &  -2.320  &  2.118e-02 {*} \tabularnewline
   &    &  C  &  {[}670.000, 1300.000)  &  -5.237e-06; {[}-6.298e-06, -4.118e-06{]}  &  -1.989  &  4.780e-02 {*} \tabularnewline
   &    &  C  &  {[}1370.000, 1580.000)  &  -5.035e-06; {[}-6.648e-06, -3.922e-06{]}  &  -2.515  &  1.256e-02 {*} \tabularnewline
   &    &  C  &  {[}1650.000, 2000.000)  &  -4.832e-06; {[}-5.820e-06, -3.681e-06{]}  &  -2.192  &  2.934e-02 {*} \tabularnewline
\bottomrule
\end{tabular}%
\end{minipage}
}
\caption{Frequency regions with statistically significant spectrum difference $d(\omega)$ between the tinnitus and control group. Uncertainty using p-values is also denoted: * ($P\le0.05$), ** ($P\le0.01$), *** ($P\le1e-3$), **** ($P\le1e-4$)}\label{tab:significant-regions}
}
\end{table}

\begin{table}\hypertarget{tab:endothelial}{%
\centering
\scalebox{0.55}{\begin{minipage}[t]{22cm}%
\begin{tabular}{ccccccccc}
\toprule
\textbf{ Frequency} & \multicolumn{2}{c}{\textbf{T3 (HbR)}} & \multicolumn{2}{c}{\textbf{T4 (HbR)}} & \multicolumn{2}{c}{\textbf{T3 (HbO)}} & \multicolumn{2}{c}{\textbf{T4 (HbO)}}\tabularnewline
\hline
\textbf{{[}mHz{]} } & t-value & p-value & t-value & p-value & t-value & p-value & t-value & p-value\tabularnewline
 3.000  &  -0.847  &  3.976e-01   &  -0.081  &  9.356e-01   &  3.275  &  1.213e-03 {*}{*}  & -0.072 & 0.9426\tabularnewline
 4.000  & -1.632  &  1.041e-01   &  0.550  &  5.827e-01   &  2.921  &  3.817e-03 {*}{*}  & -1.157 & 0.2484\tabularnewline
 5.000  &  -0.417  &  6.771e-01   &  -1.329  &  1.852e-01   &  2.289  &  2.295e-02 {*}  & -0.739 & 0.4607\tabularnewline
 6.000  &  1.660  &  9.819e-02   &  -0.731  &  4.656e-01   &  2.904  &  4.023e-03 {*}{*}  & 0.276 & 0.7824\tabularnewline
 7.000  & \textbf{ 3.032 } & \textbf{ 2.694e-03 {*}{*} } & \textbf{ 2.283 } & \textbf{ 2.333e-02 {*} } & \textbf{ 2.429 } & \textbf{ 1.587e-02 {*} } & \textbf{2.836} & \textbf{ 4.953e-03 {*}{*} }\tabularnewline
 8.000  &  4.274  &  2.763e-05 {*}{*}{*}{*}  &  0.874  &  3.829e-01   &  2.796  &  5.584e-03 {*}{*}  & 1.716 & 0.08738\tabularnewline
 9.000  &  1.494  &  1.364e-01   &  -1.155  &  2.491e-01   &  3.373  &  8.653e-04 {*}{*}{*}  & -1.54 & 0.125\tabularnewline
 10.000  &  -0.632  &  5.280e-01   &  -1.093  &  2.755e-01   &  1.233  &  2.189e-01   & -1.83 & 0.06848\tabularnewline
 11.000  &  -2.688  &  7.684e-03 {*}{*}  &  -0.577  &  5.647e-01   &  2.387  &  1.777e-02 {*}  & -0.676 & 0.4999\tabularnewline
 12.000  & \textbf{ 3.395 } & \textbf{ 8.015e-04 {*}{*}{*} } & \textbf{ 0.398 } & \textbf{ 6.911e-01  } & \textbf{ 5.136 } & \textbf{ 5.796e-07 {*}{*}{*}{*} } & \textbf{3.201} & \textbf{ 1.553e-03 {*}{*} }\tabularnewline
 13.000  &  0.555  &  5.796e-01   &  -0.225  &  8.222e-01   & \textbf{ 3.465 } & \textbf{ 6.268e-04 {*}{*}{*} } & \textbf{2.354} & \textbf{ 1.938e-02 {*} }\tabularnewline
\midrule
 14.000  &  0.596  &  5.515e-01   &  -0.307  &  7.590e-01   &  2.164  &  3.148e-02 {*}  & -1.925 & 0.05539\tabularnewline
 15.000  &  -0.642  &  5.214e-01   &  2.164  &  3.141e-02 {*}  &  0.900  &  3.691e-01   & 2.49 &  1.346e-02 {*} \tabularnewline
 15.000  &  -0.642  &  5.214e-01   &  2.164  &  3.141e-02 {*}  &  0.900  &  3.691e-01   & 2.49 &  1.346e-02 {*} \tabularnewline
 16.250  &  -1.951  &  5.221e-02   &  0.209  &  8.350e-01   &  -2.790  &  5.687e-03 {*}{*}  & -0.228 & 0.8202\tabularnewline
 17.500  &  -0.393  &  6.947e-01   &  0.190  &  8.492e-01   &  -3.616  &  3.648e-04 {*}{*}{*}  & -1.012 & 0.3125\tabularnewline
 18.750  &  0.040  &  9.678e-01   &  -2.253  &  2.515e-02 {*}  &  -2.379  &  1.816e-02 {*}  & -1.095 & 0.2744\tabularnewline
 20.000  &  -1.042  &  2.985e-01   &  -1.422  &  1.563e-01   & \textbf{ -3.159 } & \textbf{ 1.786e-03 {*}{*} } & \textbf{-1.996} & \textbf{ 4.708e-02 {*} }\tabularnewline
 21.250  & \textbf{ -2.422 } & \textbf{ 1.619e-02 {*} } & \textbf{ -3.278 } & \textbf{ 1.198e-03 {*}{*} } &  -1.656  &  9.904e-02   & -1.824 & 0.06936\tabularnewline
 22.500  & -0.260  &  7.953e-01   &  -1.131  &  2.590e-01   &  -0.516  &  6.061e-01   & -1.636 & 0.1032\tabularnewline
 23.750  &  -0.967  &  3.347e-01   &  0.637  &  5.250e-01   &  -0.690  &  4.907e-01   & 0.522 & 0.6021\tabularnewline
 25.000  &  2.771  &  6.023e-03 {*}{*}  &  -1.065  &  2.878e-01   &  0.662  &  5.087e-01   & 0.131 & 0.8958\tabularnewline
 26.250  &  2.395  &  1.736e-02 {*}  &  0.232  &  8.168e-01   & \textbf{ 2.407 } & \textbf{ 1.686e-02 {*} } & \textbf{2.838} & \textbf{ 4.929e-03 {*}{*} }\tabularnewline
 27.500  &  -0.087  &  9.309e-01   &  1.547  &  1.231e-01   &  -1.129  &  2.600e-01   & 2.162 &  3.163e-02 {*} \tabularnewline
 28.750  &  1.295  &  1.964e-01   &  -1.687  &  9.284e-02   &  0.967  &  3.343e-01   & 0.975 & 0.3305\tabularnewline
 30.000  &  0.847  &  3.977e-01   &  -0.645  &  5.198e-01   &  1.597  &  1.116e-01   & 0.624 & 0.5331\tabularnewline
 31.250  &  0.472  &  6.373e-01   &  0.702  &  4.835e-01   &  1.423  &  1.561e-01   & 2.165 &  3.134e-02 {*} \tabularnewline
 32.500  &  1.262  &  2.083e-01   &  -2.874  &  4.410e-03 {*}{*}  &  0.798  &  4.257e-01   & 2.586 &  1.031e-02 {*} \tabularnewline
 33.750  &  -0.096  &  9.238e-01   &  -2.714  &  7.124e-03 {*}{*}  &  -0.533  &  5.948e-01   & 1.257 & 0.21\tabularnewline
 35.000  &  2.380  & \textbf{ 1.811e-02 {*} } & \textbf{ -3.940 } & \textbf{ 1.069e-04 {*}{*}{*} } &  1.945  &  5.291e-02   & -0.82 & 0.413\tabularnewline
 36.250  &  1.602  &  1.104e-01   &  -2.163  &  3.151e-02 {*}  &  1.502  &  1.343e-01   & -1.882 & 0.06107\tabularnewline
 37.500  &  2.660  &  8.350e-03 {*}{*}  &  0.096  &  9.238e-01   &  3.056  &  2.496e-03 {*}{*}  & -1.308 & 0.1922\tabularnewline
 38.750  &  2.284  & \textbf{ 2.323e-02 {*} } & \textbf{ 2.609 } & \textbf{ 9.652e-03 {*}{*} } &  3.848  &  1.528e-04 {*}{*}{*}  & 1.027 & 0.3053\tabularnewline
 40.000  &  3.451  &  6.587e-04 {*}{*}{*}  &  0.245  &  8.065e-01   & \textbf{ 4.002 } & \textbf{ 8.354e-05 {*}{*}{*}{*} } & \textbf{2.01} & \textbf{ 4.558e-02 {*} }\tabularnewline
\bottomrule
\end{tabular}%
\end{minipage}
}
\caption{Very-low-frequency spectrum differences $d(\omega)$ between the tinnitus and control group. Difference uncertainties in the endothelial (3-13mHz) and neurogenic (13-40mHz) bands are denoted with their significance. Uncertainty using p-values is also denoted: * ($P\le0.05$), ** ($P\le0.01$), *** ($P\le1e-3$), **** ($P\le1e-4$)}\label{tab:endothelial}
}
\end{table}

We analyzed the inter-group contrast $d(\omega)$ for oxy- and deoxy-hemoglobin in patients with tinnitus (PT) and a healthy control (HC) group at several frequencies in the ENMRC scale \citep{WaveletOsc-Stefanovska-1999, Wavelet-Geyer-2004, PhysicsHumanCardiovascular-Stefanovska-1999}. For this procedure, data from seven subjects (Z04, Z05, Z16, and Z23 in HC, and Z07, Z09, and Z20 in the PT group) were omitted due to quality issues. We estimated the spectrum in 93 specific harmonics distributed on the overall spectrum of interest (3mHz-2Hz). Each one of the five components in the ENMRC scale was divided into segments with uniform separations of 1mHz (endothelial), 1.2mHz (neurogenic), 5mHz (myogenic), 25mHz (respiratory), 70mHz (cardiac). This division ensures that each oscillation category was split into 12-22 segments.

Remark that based on the \autoref{lem:time-frequency-uncertainty-summary}, we can expect that the rise time (time transient response) will be lower than 272.6 seconds for a minimum frequency resolution of 1mHz (using the cut-offs $\alpha=\beta=\frac 1 2$) at a sampling frequency of $f_s=20$Hz. Fast Fourier transform (FFT) requires at least $N=\frac{f_s}{\Delta\omega}=\frac{20}{0.0005}=4e5$ data points, or $2e4$ seconds, to provide a similar frequency resolution. For our analysis, we decided to configure COFRE filters with bandwidths $\tau=8.65$ such that the frequency resolution is $\Delta\omega=0.9656$mHz ($\alpha=\frac 3 4$) with a rise time $t^*=3956.62=42.8212$s ($\beta=\frac 1 4$). Our analysis was performed at a trial level, where the 18-second block were repeated ten times to compensate the filter's rise time requirement.

Moreover, we emphasize very-low-frequency intervals given the lack of tools to analyze them in finite-length samples properly. Let us consider the very-low-frequency oscillations in the spectral region of 3-20mHz, i.e., the endothelial or metabolic components in the ENMRC scale: a 5mHz-component requires 200 seconds (4000 time points) to complete a single cycle. Designing lengthy experiments that monitor the patient during several trials of this length will involve prolonged recording times that could represent an issue due to fatigue, exhaustion, discomfort, or pain during prolonged use of fNIRS devices \citep{FNIR-Kassab-2015}. Compared with the standard spectral analysis using FFT, COFRE can provide consistent estimators with limited amounts of data according to the precision required.

Using the limiting distribution of \autoref{eq:limiting-distribution}, we can estimate the mean $d(\omega)$ and construct confidence intervals to the evaluated frequencies. Across the spectrum range, it was recognized that there is a substantial variation in the spectrum contrast between HC and PT (\autoref{fig:spectrum-contrast}). As an additional consequence of \autoref{eq:limiting-distribution}, we can apply a Student's t-test two compare the PT and HC groups (\citep[eq. 1]{ImportanceNormalityAssumption-Lumley-2002}, we then determine the frequency components where the contrast is statistically significant and (p-value$<0.05$). We use the maximum p-value (maxP) procedure to combine consecutive significant harmonics (p-value$<0.05$) into significant frequency regions \citep{HypothesisSettingOrder-Song-2014, StatisticalConsiderationPsy-Wilkinson-1951}. A comprehensive list of the most distinctive intervals along with their p-values is shown in \autoref{tab:significant-regions}. Also, note that t-values can be interpreted as unitless adjusted distances between the spectrum contrast $c(\omega|g)$ in the HC and PT groups.

Furthermore, several data-driven characteristics observed in this dataset can be distinguished. First, patients with tinnitus denote a spectral contrast $c(\omega \vert g)$ higher than the healthy control group in the respiratory and cardiac components, while they seem to have a lower contrast in the 7-13mHz endothelial components, with a notorious impact in the channel T3 located in the left temporal lobe (\autoref{fig:spectrum-contrast}.C).

As expected, the magnitude of the contrasts $c(\omega \vert g)$ is inversely proportional to the frequency with a large part of the density concentrated in the low-frequency components. Therefore, to visually inspect the effects and variations between HC and PT in the entire range $3-2000$mHz, we displayed the curves of $\mathbb E c(\omega \vert g)$ and $\log \mathbb E c(\omega \vert g)$ in \autoref{fig:spectrum-contrast}. Two local maxima are remarkable in those curves : (a.) in the region containing upper-metabolic and lower-neurogenic waves, and (b.) in the cardiac region.

In the first local-maxima neighborhood, the contrast in the left hemisphere is also always lower for tinnitus subjects in the ranges 3-10mHz ($\vert\text{t-value}\vert\ge2.289$) and 11-15mHz ($\vert\text{t-value}\vert\ge2.164$) in the endothelial region. Outside this pattern, no consistent difference $d(\omega)$ was found except in the harmonics at 7mHz ($\vert\text{t-value}\vert\ge2.429$ ) and 12mHz ($\vert\text{t-value}\vert\ge0.398$ ) where the contrast in subjects tinnitus is always lower than in the control group. These types of frequencies are believed to be caused by vasomotion regulation mechanisms \citep{Vasomotion-Aalkjaer-2011, VasomotionHumanSkin-Kastrup-1989} or originated by metabolic processes in deeper layers of the brain \citep{SpontaneousLowFreq-Obrig-2000}. However, further research should be performed to assess if the observed variation is solely related to the tinnitus condition and not from tinnitus' side effects or symptoms.

The second local maxima are observed in the cardiac region. Note that the contrast response of the tinnitus group in this entire interval is always higher than the control group in every analyzed condition: left and right hemisphere, and oxy- and deoxy-hemoglobin. We can understand from this metric that sound is correlated with a high variation of cardiovascular oscillations in the subjects with tinnitus compared to the healthy group. The dataset used in this study did not include an intraarterial subtraction angiography to evaluate subjective pulsatile tinnitus as Sila et al.~recommended \citep{PulsatileTinnitus-Sila-1987}. However, we can postulate that the observed variations in cardiac components could be biomarkers of pulsatile tinnitus in the current sample dataset.

Even though that distinctive spectrum patterns were found between PT and HC, no common pattern was found in the respiratory waves across chromophores and hemispheres. Nevertheless, we found that some neurogenic-myogenic signatures appear in the range of 37-50mHz with $\vert\text{t-values}\vert$ between $2.010$ (HbO, T4) and $3.046$ (HbO, T3).

\section{Conclusion}

This paper proposes a new method for spectrum estimation, the Complex-Pole Filter Representation (COFRE), to accurately discover discriminatory frequency signatures on fNIRS signals between patients with tinnitus and healthy control groups as a proof-of-concept. The spectral information contained in biomedical signals can be modeled using filter banks with narrow bandwidths. COFRE used this representation as a framework to propose a spectrum estimation based on narrow-band filters with a single complex pole (\autoref{eq:car}).

COFRE inherits several relevant characteristics from the narrow-band filters that comprised it. COFRE filters can be configured to have the desired frequency resolution (\autoref{lem:frequency-resolution}) or maximum rise time as a transient time response metric (\autoref{lem:rise-time}). As expected, in this setting, improving the time response could reduce the frequency response and vice versa. An upper-bound for this joint restriction was also formalized in \autoref{lem:time-frequency-uncertainty}. Furthermore, COFRE filters have a constant time and space complexity $\mathcal{O}(1)$ that support real-time applications or processing on memory-constrained systems.

We benefit from the properties of the COFRE spectrum representation to examine notable changes in the hemodynamic characteristics that could be related to a tinnitus condition. The interpretation was performed within the biological framework described by the ENMRC spectral division. Thus, we could observe specific oscillations that could serve as biomarkers: at 7mHz in the metabolic frequency region; and variations in the 30-50mHz in the neurogenic/myogenic band. Furthermore, notable, statistically significant, differences were denoted in the whole spectral bands related to respiratory and cardiac responses.

Even though this research was based on hemodynamic responses in tinnitus, we believe COFRE has the ability to be used in other studies involving biomedical signals where (a.) accurate estimation of exact frequency components is needed, (b.) spectral information is biologically interpretable, or (c.) time- or space-optimal algorithms for spectral estimation are expected.

\clearpage
\begin{appendices}

\section{Narrow band-pass filter properties}

\subsection{Symmetric response}\label{sec:symmetric-response}

\begin{lemma*}[Symmetry response]\label{lem:conditions-proof}A complex single-pole IIR filter defined by the transfer function $M\left(\omega;\rho,\omega^{*}\right)$ is symmetric around $\omega^{*}$, i.e, $M\left(\omega;\rho,\omega^{*}-\Delta\omega\right)=M\left(\omega;\rho,\omega^{*}+\Delta\omega\right)\,\Delta \omega\in[0,\frac 1 2]$.\end{lemma*}

\begin{proof}The magnitude response (\autoref{eq:magnitude}) can also be expressed as%
\begin{equation}%
{
M\left(z;\rho,\omega^{*}\right)
=\left|H\left(\exp\left(j2\pi\omega\right)\right)\right|
=\left(\left(1-\rho\cos\left(2\pi\left(\omega^{*}-\omega\right)\right)\right)^{2}+\rho^{2}\sin^{2}\left(2\pi\left(\omega^{*}-\omega\right)\right)\right)^{-\frac{1}{2}}
}%
\end{equation}%
Now, let us examine the magnitude in the neighborhood of $\omega^{*}$ through $\omega=\omega^{*}+D$, $D>0$:%
\begin{equation}%
{ 
\left|H\left(\exp\left(j2\pi\left(\omega^{*}+D\right)\right)\right)\right|
= \left(\left(1-\rho\cos\left(2\pi D\right)\right)^{2}+\rho^{2}\sin^{2}\left(2\pi D\right)\right)^{-\frac{1}{2}}
}%
\end{equation}%

Then, it is straightforward to corroborate that%
\begin{equation}%
{
\left|H\left(\exp\left(j2\pi\left(\omega^{*}-D\right)\right)\right)\right|=\left(\left(1-\rho\cos\left(2\pi D\right)\right)^{2}+\rho^{2}\sin^{2}\left(-2\pi D\right)\right)^{-\frac{1}{2}}=\left|H\left(\exp\left(j2\pi\left(\omega^{*}+D\right)\right)\right)\right|
}%
\end{equation}%
\end{proof}

\subsection{Unique maximum}\label{sec:unique-maximum}

\begin{lemma*}[Unique maximum]\label{lem:uniqueness-proof}The filter $M\left(\omega;\rho,\Delta\omega\right)$ has a single and unique maximum located at $\omega^{*}$ and%
\begin{equation}%
{
\max M\left(\omega;\rho,\omega\right)=M\left(\omega;\rho,\omega^{*}\right)=\left|1-\rho\right|^{-1}
}%
\end{equation}%
\end{lemma*}

\begin{proof}Let us examine the maximum filter magnitude (\autoref{eq:magnitude}),%
\begin{equation}%
{
\begin{aligned}
\max M\left(\omega;\rho,\omega\right)
= \max\left|H\left(e^{j2\pi\omega}\right)\right|
  & = \max\frac{1}{\left|
    1-\rho e^{j2\pi\left(\omega^{*}-\omega\right)}
  \right|}
= \min\left|
    1-\rho e^{j2\pi\left(\omega^{*}-\omega\right)}
  \right| \\
  & = \min\sqrt{
    \left(1-\rho\cos\left(2\pi\left(\omega^{*}-\omega\right)\right)\right)^{2}
    + \rho^{2}\sin^{2}\left(2\pi\left(\omega^{*}-\omega\right)\right)
  }
\end{aligned}
}%
\end{equation}%
Recall that $f(z)=\sqrt{z}$ is a concave function. Therefore, $\min \sqrt{z} = \sqrt{\min z}$ and%
\begin{equation}%
{
\max\left|H\left(e^{j2\pi\omega}\right)\right|
  = \sqrt{\min f\left(\omega\right)}
}%
\end{equation}%
where $f\left(\omega\right) = \left(1-\rho\cos\left(2\pi\omega^{*}-2\pi\omega\right)\right)^{2} + \rho^{2}\sin^{2}\left(2\pi\omega^{*}-2\pi\omega\right)$.

Note that the extrema in $f\left(\omega\right)$ are going to satisfy

\begin{equation}%
{
\begin{aligned}
\frac{d}{d\omega}f\left(\omega\right)
  & =-2\rho\sin\left(2\pi\left(\omega^{*}-\omega\right)\right)
    = 0
\end{aligned}
}%
\end{equation}%

As a consequence of this expression, we have a periodic number of solutions (extreme values) at $\omega=\omega^{*}+\frac{1}{2}\ell,\,\ell=0,1,\ldots$. However, in the frequency interval defined by the filter $H\left(e^{j\omega}\right)$: $\omega\in\left[0,\frac 1 2\right]$, the only critical point is located at $\omega=\omega^{*}$.

To corroborate that the extreme at $\omega^*$ is a maximum, we can evaluate the second derivative of $f\left(\omega\right)$ at the point $\omega=\omega^{*}$:

\begin{equation}%
{
\begin{aligned}
g\left(\omega^*\right)
  & =
   \left.
     \frac{d^{2}}
          {d\omega^{2}}f\left(\omega\right)
   \right\vert_{\omega=\omega^*}
  & = 2\rho\cos\left(2\pi\omega^{*}-2\pi\omega\right)
    = 2\rho > 0
\end{aligned}
}%
\end{equation}%
Then, $g\left(\omega^*\right)$ is always positive for any $0<\rho\le1$, and $\omega=\omega^{*}$ is the unique maximum in $\left[0,\pi\right]$.
\end{proof}

\subsection{Frequency resolution}\label{sec:frequency-resolution}

\begin{lemma*}[Frequency resolution configuration]\label{lem:frequency-resolution-proof}The cosine of the frequency resolution of a filter $M\left(\omega;\rho,\omega^{*}\right)$ is a second-order rational function of the complex autoregressive modulus $\rho$:%
\begin{equation}%
{
\begin{aligned}
\Delta\omega
&= \frac{1}{2\pi}
   \arccos
   \left(\frac{1}{2\rho}
  \left(\rho^{2}+1-\alpha^{-2}\left(1-\rho\right)^{2}\right)
   \right)
\end{aligned}
}%
\end{equation}%
\end{lemma*}

\begin{proof}Recall the magnitude response (\autoref{eq:magnitude}) at a frequency $\omega$:%
\begin{equation}%
{
\begin{aligned}
\left|H\left(e^{j\omega}\right)\right|
  & =  \frac{1}
        {\left|
           1-\rho e^{j2\pi\left(\omega^{*}-\omega\right)}
        \right|}
  &  = \frac
   {\left|
       e^{j2\pi\omega}
   \right|}
   {\left|
       e^{j2\pi\omega}-\rho e^{j2\pi\omega^{*}}
   \right|}
  & = \frac
   {1}
   {d\left(\omega\right)}
\end{aligned}
}%
\end{equation}%
where $d\left(\omega\right) =\left|e^{j2\pi\omega}-\rho e^{j2\pi\omega^{*}}\right|$.

\begin{figure}
\hypertarget{fig:pole-distances}{%
\centering
\includegraphics[width=0.35\textwidth,height=\textheight]{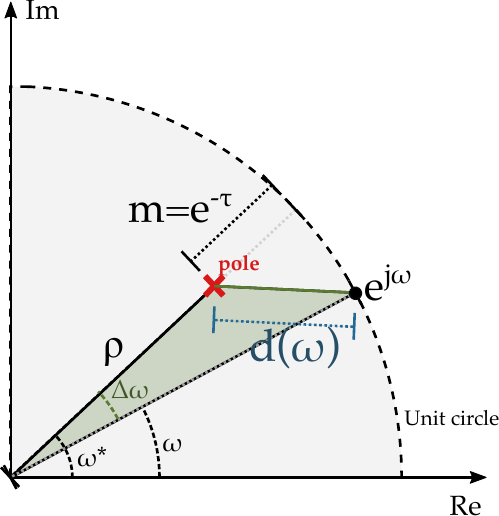}
\caption{Geometric interpretation of the complex pole of a filter $M\left(\omega;\rho,\omega^*\right)$}\label{fig:pole-distances}
}
\end{figure}

Note that $d(\omega)$ can be interpreted as the distance between the complex poles at $\omega$ and $\omega^*$. In consequence, $d(\omega^*)=\vert1-\rho\vert$. A graphical representation of $d\left(\omega\right)$ is depicted in \autoref{fig:pole-distances}.

Let $\omega=\omega^*+\Delta\omega$, where $\Delta\omega$ is the frequency resolution (\autoref{def:alternative-frequency-resolution}), then%
\begin{equation}%
{
d\left(\omega\right)
  = \frac{1}{M\left(\omega;\rho,\omega^*+\Delta\omega\right)}
  = \frac{1}{\alpha M\left(\omega;\rho,\omega^*\right)}
  = \frac{1}{\alpha} d\left(\omega^*\right)
  = \frac{1}{\alpha} \left\vert 1 - \rho \right\vert
}%
\end{equation}%

From the geometrical interpretation (\autoref{fig:pole-distances}), it is straightforward to infer that $\Delta\omega$ and $\rho$ are related through the law of cosines:%
\begin{equation}%
{
2\rho\cos2\pi\Delta\omega
  = \rho^{2}+1-d^{2}\left(\omega\right)
  = \rho^{2}+1-\alpha^{-2}\left(1-\rho\right)^{2}
}\label{eq:law-cos}%
\end{equation}%
\end{proof}

\subsection{Frequency resolution}\label{sec:frequency-optimal-bandwidth}

\begin{corollary*}[Frequency-optimal bandwidth]\label{lem:frequency-optimal-bandwidth-proof}The minimum bandwidth $\tau$ to ensure a frequency resolution $\Delta\omega$ under a cut-off $\alpha$ is given by\end{corollary*}

\begin{equation}%
{
\begin{aligned}
\tau
  &= -\log\left(
        1
        - \frac{\cos2\pi\Delta\omega-\alpha^{-2}}{1-\alpha^{-2}}
        - \frac{1}{1-\alpha^{-2}}
           \sqrt{\cos^{2}2\pi\Delta\omega-\left(1-\alpha^{-2}\right)^{2}}
      \right)
\end{aligned}
}%
\end{equation}%

\begin{proof}From the relationship between $\Delta\omega$ and $\rho$ denoted in \autoref{eq:law-cos}, we obtain%
\begin{equation}%
{
0=\rho^{2}\left(1-\alpha^{-2}\right)+2\rho\left(\alpha^{-2}-\cos2\pi\Delta\omega\right)+\left(1-\alpha^{-2}\right)
}%
\end{equation}%
Therefore,%
\begin{equation}%
{
\rho=\frac{\cos2\pi\Delta\omega-\alpha^{-2}}{1-\alpha^{-2}}\pm\frac{1}{1-\alpha^{-2}}\sqrt{\cos^{2}2\pi\Delta\omega-\left(1-\alpha^{-2}\right)^{2}}
}%
\end{equation}%
Note that given that $\rho\ge0$, the negative root of the quadratic expression will be ignored.

Finally, recall the equivalency between the bandwidth $\tau$ and the modulus $\rho$ : $\rho=1-e^{-\tau}$. Then,%
\begin{equation}%
{
e^{-\tau}=1-\frac{\cos2\pi\Delta\omega-\alpha^{-2}}{1-\alpha^{-2}}-\frac{1}{1-\alpha^{-2}}\sqrt{\cos^{2}2\pi\Delta\omega-\left(1-\alpha^{-2}\right)^{2}}
}%
\end{equation}%
\end{proof}

\subsection{Rise time}\label{sec:rise-time}

\begin{lemma*}[Rise time configuration]\label{lem:rise-time-proof}The rise time of a filter $M\left(\omega;\rho,\omega^{*}\right)$ is inversely proportional to the logarithm of $\rho$ and proportional to the logarithm of the complement of the cut-off factor $\beta$:%
\begin{equation}%
{
t^{*}=\log_{\rho}\left(\frac{1-\beta}{\rho}\right)=\frac{\log\left(1-\beta\right)}{\log\left(\rho\right)}-1
}%
\end{equation}%
\end{lemma*}

\begin{proof}Recall the transfer function of the filter (\autoref{eq:filter-def-z}):%
\begin{equation}%
{
H\left(z\right)=\frac{1}{1-\rho e^{j2\pi\omega^{*}}z^{-1}}=F\left(z;\rho,\omega^{*}\right)
}%
\end{equation}%

Without loss of generality, we can assume a complex input signal $x\left(t\right)=\exp\left(j2\pi vt\right)$, where $v\in\left[0,\frac 1 2\right]$. Then, the output of the filter is%
\begin{equation}%
{
Y\left(z\right)=\frac{1}{1-\rho e^{j2\pi\omega^{*}}z^{-1}}\,\frac{1}{1-e^{j2\pi v}z^{-1}}
}%
\end{equation}%

Solving by partial fractions, we obtain%
\begin{equation}%
{
Y\left(z\right)=\frac{1}{e^{j2\pi v}-\rho e^{j2\pi\omega^{*}}}\,\left(\frac{e^{j2\pi v}}{1-e^{j2\pi v}z^{-1}}-\frac{\rho e^{j2\pi\omega^{*}}}{1-\rho e^{j2\pi\omega^{*}}z^{-1}}\right)
}%
\end{equation}%

Therefore, the inverse z-transform of $Y(z)$: $y(t)$ is given by%
\begin{equation}%
{
y\left(t\right)=\frac{1}{e^{j2\pi v}-\rho e^{j2\pi\omega^{*}}}\,\left(e^{j2\pi v}e^{j2\pi vt}-\rho^{t+1}e^{j2\pi\omega^{*}}e^{j2\pi \omega^{*}t}\right)
}%
\end{equation}%

Now, let us assume that the input signal is resonating at the central frequency of the filter ($v=\omega^{*}$):%
\begin{equation}%
{
y\left(t\right)
  =\frac{1-\rho^{t+1}}{1-\rho}\,e^{j2\pi\omega^{*}t}
}%
\end{equation}%

Then, the real component of $y(t)$ is defined as%
\begin{equation}%
{
y^{*}\left(t\right)
  = \text{Re}\left(y\left(t\right)\right)
  = \frac{1-\rho^{t+1}}{1-\rho}\,\cos\left(2\pi\omega^{*}t\right)
  = A\left(t\right)\cos\left(2\pi\omega^{*}t\right)
}%
\end{equation}%

We can denote that the envelope $A\left(t\right)$ is determining the transient behavior of the filter. In a stable filter ($\rho<1$), the envelope converges to $A_{\max}$ at the infinite:%
\begin{equation}%
{
\lim_{t\rightarrow\infty}
\left(\frac{1-\rho^{t+1}}{1-\rho}\right)
  = \frac{1}{1-\rho}=A_{\max}
}%
\end{equation}%

Consequently, the time $t^{*}$ needed to for $A\left(t\right)$ to ``rise'' from $0$ to $\beta A_{\max}$ (\autoref{def:rise-time}) satisfies%
\begin{equation}%
{
\beta\frac{1}{1-\rho}
  = \frac{1-\rho^{t^{*}+1}}
         {1-\rho}
}%
\end{equation}%

Therefore, $t^{*}$, when $0<\rho<1$, is determined by%
\begin{equation}%
{
t^{*}
  = \frac{\log\left(1-\beta\right)}
        {\log\left(\rho\right)}-1
  = \log_{\rho}\left(\frac{1-\beta}{\rho}\right)
}\label{eq:rise-time-rho}%
\end{equation}%
\end{proof}

\subsection{Time-optimal bandwidth}\label{sec:rise-optimal-bandwidth}

\begin{corollary*}[Time-optimal bandwidth]\label{lem:rise-optimal-bandwidth-proof}The minimum bandwidth $\tau$ to ensure a rise time $t^{*}$ (under a cut-off $\beta$) is%
\begin{equation}%
{
\tau=-\log\left(1-\left(1-\beta\right)^{\frac{1}{1+t^{*}}}\right)
}%
\end{equation}%
\end{corollary*}

\begin{proof}Recall \autoref{eq:rise-time-rho}, and rearrange it:%
\begin{equation}%
{
\log\left(\rho\right)=\frac{\log\left(1-\beta\right)}{1+t^{*}}
}%
\end{equation}%

Then, we can describe $\rho$ as a function of the cut-off and rise-time $t^*$:%
\begin{equation}%
{
\rho=\left(1-\beta\right)^{\frac{1}{1+t^{*}}}
}%
\end{equation}%

Now, given the bandwidth-modulus relationship $\rho=1-e^{-\tau}$,%
\begin{equation}%
{
e^{-\tau}=1-\left(1-\beta\right)^{\frac{1}{1+t^{*}}}
}%
\end{equation}%
\end{proof}

\subsection{Time-frequency constraint}\label{sec:time-frequency-uncertainty}

\begin{lemma*}[Joint time-frequency constraint]\label{lem:time-frequency-uncertainty-proof}Given the cut-offs $\alpha$ and $\beta$, the rise time $t^{*}$ and the frequency resolution $\Delta\omega$ are mutually constrained by%
\begin{equation}%
{
t^{*}\Delta\omega
   < -\log\left(
    1-\beta\right)
 \left(\frac{\sqrt{\left(\alpha\pi\Delta\omega\right)^{2}-\alpha^{2}+1}}{2\pi\alpha}+\frac{\Delta\omega}{2}\right)-\Delta\omega
}%
\end{equation}%
\end{lemma*}

\begin{proof}Let us analyze the Maclaurin series of the cosine function:%
\begin{equation}%
{
\cos z=\sum_{k=0}^{\infty}\frac{1}{\left(2k\right)!}\left(-1\right)^{k}z^{2k}=1-\frac{1}{2!}z^{2}+\frac{1}{4!}z^{4}-\frac{1}{6!}z^{6}+\cdots
}%
\end{equation}%

The radius of convergence $r_{k\mapsto k+1}^{(c)}$ of this series is given by%
\begin{equation}%
{
r_{k\mapsto k+1}^{(c)}=\frac{\frac{1}{\left(2k+2\right)!}\left(-1\right)^{k+1}z^{2\left(k+1\right)}}{\frac{1}{\left(2k\right)!}\left(-1\right)^{k}z^{2k}}=-\frac{z^{2}}{\left(2k+2\right)\left(2k+1\right)}
}%
\end{equation}%

It is remarkable that $\left|r_{k\mapsto k+1}^{(c)}\right|<1$, and consequently,%
\begin{equation}%
{
\left(1-\left|r_{k\mapsto k+1}^{(c)}\right|\left(1-\left|r_{k+1\mapsto k+2}^{(c)}\right|\left(1-\cdots\right)\right)\right)<1
}%
\end{equation}%

Now, let us examine the series expansion of the logarithm function,%
\begin{equation}%
{
\begin{aligned}
\log\left(1-z\right) & =-\sum_{k=1}^{\infty}\frac{1}{k}z^{k}=-z-\frac{1}{2}z^{2}-\frac{1}{3}z^{3}-\frac{1}{4}z^{4}+\cdots
\end{aligned}
}%
\end{equation}%
with its inherent radius of convergence,%
\begin{equation}%
{
r_{k\mapsto k+1}^{(l)}=\frac{\frac{-1}{\left(k+1\right)}z^{\left(k+1\right)}}{\frac{-1}{k}z^{k}}=\frac{1}{1+\frac{1}{k}}z
}%
\end{equation}%

It is clear that the convergence of the series is ensured when $z<1$, satisfying also the following inequality:%
\begin{equation}%
{
\left(1+r_{k\mapsto k+1}^{(l)}\left(1+r_{k+1\mapsto k+2}^{(l)}\left(1+\cdots\right)\right)\right)>1
}%
\end{equation}%

Provided the previous inequalities, we can establish some boundaries for $\cos z$ and $\log\left(1+z\right)$:%
\begin{equation}%
{
\cos z=1-\frac{z^{2}}{2}\left(1-\frac{z^{2}}{12}\left(1-\frac{z^{2}}{40}\left(1-\cdots\right)\right)\right)>1-\frac{z^{2}}{2}
}\label{eq:bound-cos}%
\end{equation}%
\begin{equation}%
{
\log\left(1-z\right)=-z\left(1+\frac{z}{2}\left(1+\frac{z}{1+\frac{1}{2}}\left(1-\cdots\right)\right)\right)<-z\quad0<z<1
}\label{eq:bound-log}%
\end{equation}%

Now, recall \autoref{eq:law-cos},%
\begin{equation}%
{
\cos2\pi\Delta\omega=\frac{1}{2\rho}\left(1+\rho^{2}-\alpha^{-2}\left(1-\rho\right)^{2}\right)
}%
\end{equation}%

By applying the upperbound of \autoref{eq:bound-cos}, we obtain%
\begin{equation}%
{
1-\frac{1}{2}\left(2\pi\Delta\omega\right)^{2}>\frac{1}{2\rho}\left(1+\rho^{2}-\alpha^{-2}\left(1-\rho\right)^{2}\right)
}%
\end{equation}%

For brevity in the notation, let $v=2\pi\Delta\omega$ and $u=\left(1-\rho\right)$:%
\begin{equation}%
{
\begin{aligned}
v^{2} 
   & < 2-\frac{1}{\rho}
         \left(1+\rho^{2}-\alpha^{-2}u^{2}\right)\\
   & < -\frac{1}{\rho}
         \left(1-2\rho+\rho^{2}-\alpha^{-2}u^{2}\right)\\
   & <-\frac{1}{1-u}
          \left(u^{2}-\alpha^{-2}u^{2}\right)
\end{aligned}
}%
\end{equation}%

Furthermore, let $a=\left(1-\alpha^{-2}\right)$, then%
\begin{equation}%
{
w^{2}<-a\frac{u^{2}}{1-u}
}%
\end{equation}%
\begin{equation}%
{
au^{2}-w^{2}u+w^{2}<0
}\label{eq:ineq-w}%
\end{equation}%

Later, recall \autoref{eq:rise-time-rho},%
\begin{equation}%
{
\log\left(1-\left(1-\rho\right)\right)=\frac{1}{\left(1+t^{*}\right)}\log\left(1-\beta\right)
}%
\end{equation}%

Let $b=\log\left(1-\beta\right)$ and $T=\left(1+t^{*}\right)$, and use the boundary in \autoref{eq:bound-log},%
\begin{equation}%
{
\begin{aligned}
-u & <\frac{1}{\left(1+t^{*}\right)}\log\left(1-\beta\right)=\frac{b}{T}\\
u & >-\frac{b}{T}
\end{aligned}
}%
\end{equation}%

Note that $u\in(0,1)$ and $a<0$, then%
\begin{equation}%
{
-wu<w\frac{b}{T}
}%
\end{equation}%
\begin{equation}%
{
au^{2}<a\frac{b^{2}}{T^{2}}
}%
\end{equation}%

Replacing both expressions in \autoref{eq:ineq-w},%
\begin{equation}%
{
a\frac{b^{2}}{T^{2}}-w\frac{b}{T}+w^{2}<0
}%
\end{equation}%

Therefore,%
\begin{equation}%
{
\left(Tw\right)^{2}+b\left(Tw\right)w+ab^{2}<0
}%
\end{equation}%

Factorizing the left side of the inequality,%
\begin{equation}%
{
\left(Tw-\frac{1}{2}b\left(\sqrt{w^{2}-4a}+w\right)\right)\left(Tw-\frac{1}{2}b\left(-\sqrt{w^{2}-4a}+w\right)\right)<0
}%
\end{equation}%
\begin{equation}%
{
Tw<\frac{1}{2}b\left(\sqrt{w^{2}-4a}+w\right)
}%
\end{equation}%

Now, reformulating the inequality in terms of $\Delta\omega$ and $t^{*}$,%
\begin{equation}%
{
t^{*}\Delta\omega
   < -\log\left(
        1-\beta\right)
     \left(\frac{\sqrt{\left(\alpha\pi\Delta\omega\right)^{2}-\alpha^{2}+1}}{2\pi\alpha}+\frac{\Delta\omega}{2}\right)-\Delta\omega
}\label{eq:time-frequency-base}%
\end{equation}%
Note that $\left(\alpha\pi\Delta\omega\right)^{2}-\alpha^{2}+1>0\quad\forall\Delta\omega\in\left[0,\frac{1}{2}\right]$.

Let us consider bandpass filters with frequency resolution $\Delta\omega<\frac{1}{5}$, then the time-frequency constraint can be upper-bounded by%
\begin{equation}%
{
t^{*}\Delta\omega<-\log\left(1-\beta\right)\left(\frac{\sqrt{1-\left(1+\frac{4}{25}\pi^{2}\right)\alpha^{2}}}{2\pi\alpha}+\frac{1}{10}\right)-\frac{1}{5}
}%
\end{equation}%

Furthermore, in filters with extremely low frequency resolution $\Delta\omega\rightarrow0$, \autoref{eq:time-frequency-base} can be simplified to%
\begin{equation}%
{
t^{*}\Delta\omega<-\log\left(1-\beta\right)\left(\frac{\sqrt{1-\alpha^{2}}}{2\pi\alpha}\right)
}%
\end{equation}%
\end{proof}

\bibliographystyle{elsarticle-num}
\bibliography{Library,ExtraBib}

\begin{thebibliography}{10}
\expandafter\ifx\csname url\endcsname\relax
  \def\url#1{\texttt{#1}}\fi
\expandafter\ifx\csname urlprefix\endcsname\relax\def\urlprefix{URL }\fi
\expandafter\ifx\csname href\endcsname\relax
  \def\href#1#2{#2} \def\path#1{#1}\fi

\bibitem{TextbookTinnitus-Moller-2011}
A.~R. M{\o}ller, B.~Langguth, D.~De~Ridder, T.~Kleinjung (Eds.), Textbook of
  {{Tinnitus}}, {Springer New York}, {New York, NY}, 2011.
\newblock \href {https://doi.org/10.1007/978-1-60761-145-5}
  {\path{doi:10.1007/978-1-60761-145-5}}.

\bibitem{TinnitusTreatmentClinical-Tyler-2006}
R.~S. Tyler (Ed.), Tinnitus Treatment: Clinical Protocols, {Thieme}, {New
  York}, 2006.

\bibitem{NeuroPhysApproachTinnitus-Jastreboff-1996}
P.~J. Jastreboff, W.~C. Gray, S.~L. Gold, Neurophysiological approach to
  tinnitus patients, The American Journal of Otology 17~(2) (1996) 236--240.

\bibitem{PhantomAuditoryPerception-Jastreboff-1990}
P.~J. Jastreboff, Phantom auditory perception (tinnitus): Mechanisms of
  generation and perception, Neuroscience Research 8~(4) (1990) 221--254.
\newblock \href {https://doi.org/10.1016/0168-0102(90)90031-9}
  {\path{doi:10.1016/0168-0102(90)90031-9}}.

\bibitem{TinnitusMultidisciplinaryApproach-Baguley-2013}
D.~Baguley (Ed.), Tinnitus: A Multidisciplinary Approach, 2nd Edition,
  {Wiley-Blackwell}, {Chichester, West Sussex, UK ; Hoboken, NJ, USA}, 2013.

\bibitem{NeuroscienceTinnitus-Eggermont-2012}
J.~J. Eggermont, The Neuroscience of Tinnitus, {Oxford University Press},
  {Oxford, United Kingdom}, 2012.

\bibitem{NeuroimagingNIR-Sevy-2010}
A.~B.~G. Sevy, H.~Bortfeld, T.~J. Huppert, M.~S. Beauchamp, R.~E. Tonini, J.~S.
  Oghalai, Neuroimaging with near-infrared spectroscopy demonstrates
  speech-evoked activity in the auditory cortex of deaf children following
  cochlear implantation, Hearing Research 270~(1-2) (2010) 39--47.
\newblock \href {https://doi.org/10.1016/j.heares.2010.09.010}
  {\path{doi:10.1016/j.heares.2010.09.010}}.

\bibitem{CorticalActivationPatterns-Olds-2016}
C.~Olds, L.~Pollonini, H.~Abaya, J.~Larky, M.~Loy, H.~Bortfeld, M.~S.
  Beauchamp, J.~S. Oghalai, Cortical {{Activation Patterns Correlate}} with
  {{Speech Understanding After Cochlear Implantation}}, Ear \& Hearing 37~(3)
  (2016) e160--e172.
\newblock \href {https://doi.org/10.1097/AUD.0000000000000258}
  {\path{doi:10.1097/AUD.0000000000000258}}.

\bibitem{FNIR-Schecklmann-2014}
M.~Schecklmann, A.~Giani, S.~Tupak, B.~Langguth, V.~Raab, T.~Polak,
  C.~V{\'a}rallyay, W.~Harnisch, M.~J. Herrmann, A.~J. Fallgatter, Functional
  {{Near}}-{{Infrared Spectroscopy}} to {{Probe State}}- and {{Trait}}-{{Like
  Conditions}} in {{Chronic Tinnitus}}: {{A Proof}}-of-{{Principle Study}},
  Neural Plasticity 2014 (2014) 1--8.
\newblock \href {https://doi.org/10.1155/2014/894203}
  {\path{doi:10.1155/2014/894203}}.

\bibitem{HumanAuditoryAdjacent-Issa-2016}
M.~Issa, S.~Bisconti, I.~Kovelman, P.~Kileny, G.~J. Basura, Human {{Auditory}}
  and {{Adjacent Nonauditory Cerebral Cortices Are Hypermetabolic}} in
  {{Tinnitus}} as {{Measured}} by {{Functional Near}}-{{Infrared Spectroscopy}}
  ({{fNIRS}}), Neural Plasticity 2016 (2016) 1--13.
\newblock \href {https://doi.org/10.1155/2016/7453149}
  {\path{doi:10.1155/2016/7453149}}.

\bibitem{WaveletOsc-Stefanovska-1999}
A.~Stefanovska, M.~Bracic, H.~Kvernmo, Wavelet analysis of oscillations in the
  peripheral blood circulation measured by laser {{Doppler}} technique, IEEE
  Transactions on Biomedical Engineering 46~(10) (Oct./1999) 1230--1239.
\newblock \href {https://doi.org/10.1109/10.790500}
  {\path{doi:10.1109/10.790500}}.

\bibitem{Wavelet-Geyer-2004}
M.~J. Geyer, Y.-K. Jan, D.~M. Brienza, M.~L. Boninger, Using wavelet analysis
  to characterize the thermoregulatory mechanisms of sacral skin blood flow,
  The Journal of Rehabilitation Research and Development 41~(6) (2004) 797.
\newblock \href {https://doi.org/10.1682/JRRD.2003.10.0159}
  {\path{doi:10.1682/JRRD.2003.10.0159}}.

\bibitem{PhysicsHumanCardiovascular-Stefanovska-1999}
A.~Stefanovska, Physics of the human cardiovascular system, Contemporary
  Physics 40~(1) (1999) 31--55.
\newblock \href {https://doi.org/10.1080/001075199181693}
  {\path{doi:10.1080/001075199181693}}.

\bibitem{VasomotionHumanSkin-Kastrup-1989}
J.~Kastrup, J.~B{\"u}low, N.~A. Lassen, Vasomotion in human skin before and
  after local heating recorded with laser {{Doppler}} flowmetry. {{A}} method
  for induction of vasomotion, International Journal of Microcirculation,
  Clinical and Experimental 8~(2) (1989) 205--215.

\bibitem{InvolvementSympatheticNerve-Soderstrom-2003}
T.~S{\"o}derstr{\"o}m, A.~Stefanovska, M.~Veber, H.~Svensson, Involvement of
  sympathetic nerve activity in skin blood flow oscillations in humans,
  American Journal of Physiology-Heart and Circulatory Physiology 284~(5)
  (2003) H1638--H1646.
\newblock \href {https://doi.org/10.1152/ajpheart.00826.2000}
  {\path{doi:10.1152/ajpheart.00826.2000}}.

\bibitem{DynamicTrackingMicrovascular-Mendelson-2020}
A.~A. Mendelson, A.~Rajaram, D.~Bainbridge, K.~S. Lawrence, T.~Bentall,
  M.~Sharpe, M.~Diop, C.~G. Ellis, Dynamic tracking of microvascular hemoglobin
  content for continuous perfusion monitoring in the intensive care unit: Pilot
  feasibility study, Journal of Clinical Monitoring and Computing (Oct. 2020).
\newblock \href {https://doi.org/10.1007/s10877-020-00611-x}
  {\path{doi:10.1007/s10877-020-00611-x}}.

\bibitem{DifferencesNetInformation-Urquhart-2020}
E.~L. Urquhart, X.~Wang, H.~Liu, P.~J. Fadel, G.~Alexandrakis, Differences in
  {{Net Information Flow}} and {{Dynamic Connectivity Metrics Between
  Physically Active}} and {{Inactive Subjects Measured}} by {{Functional
  Near}}-{{Infrared Spectroscopy}} ({{fNIRS}}) {{During}} a {{Fatiguing
  Handgrip Task}}, Frontiers in Neuroscience 14 (2020) 167.
\newblock \href {https://doi.org/10.3389/fnins.2020.00167}
  {\path{doi:10.3389/fnins.2020.00167}}.

\bibitem{SimulFNIRSThermal-Pinti-2015}
P.~Pinti, D.~Cardone, A.~Merla, Simultaneous {{fNIRS}} and thermal infrared
  imaging during cognitive task reveal autonomic correlates of prefrontal
  cortex activity, Scientific Reports 5~(1) (2015) 17471.
\newblock \href {https://doi.org/10.1038/srep17471}
  {\path{doi:10.1038/srep17471}}.

\bibitem{EffectCerebralVasomotion-Bosch-2017}
B.~M. Bosch, A.~Bringard, G.~Ferretti, S.~Schwartz, K.~Igl{\'o}i, Effect of
  cerebral vasomotion during physical exercise on associative memory, a
  near-infrared spectroscopy study, Neurophotonics 4~(4) (2017) 041404.
\newblock \href {https://doi.org/10.1117/1.nph.4.4.041404}
  {\path{doi:10.1117/1.nph.4.4.041404}}.

\bibitem{PredominantEndothelialVasomotor-Zhang-2014}
Z.~Zhang, R.~Khatami, Predominant endothelial vasomotor activity during human
  sleep: A near-infrared spectroscopy study, European Journal of Neuroscience
  40~(9) (2014) 3396--3404.
\newblock \href {https://doi.org/10.1111/ejn.12702}
  {\path{doi:10.1111/ejn.12702}}.

\bibitem{StatisticalDigitalSig-Hayes-1996}
M.~H. Hayes, Statistical Digital Signal Processing and Modeling, {John Wiley
  and Sons}, {New York}, 1996.

\bibitem{QuantifyingLowFreqFluctuations-Folgosi-2011}
M.~S. Folgosi, G.~E.~C. Nogueira, Quantifying low-frequency fluctuations in the
  laser {{Doppler}} flow signal from human skin, in: {{SPIE BiOS}}, {San
  Francisco, California, USA}, 2011, p. 789811.
\newblock \href {https://doi.org/10.1117/12.874080}
  {\path{doi:10.1117/12.874080}}.

\bibitem{RecognizingBrainActivities-Khoa-2008}
T.~Q.~D. Khoa, M.~Nakagawa, Recognizing brain activities by functional
  near-infrared spectroscope signal analysis, Nonlinear Biomedical Physics
  2~(1) (2008) 3.
\newblock \href {https://doi.org/10.1186/1753-4631-2-3}
  {\path{doi:10.1186/1753-4631-2-3}}.

\bibitem{HandbookBrainTheory-Arbib-2003}
M.~A. Arbib (Ed.), The Handbook of Brain Theory and Neural Networks, 2nd
  Edition, {MIT Press}, {Cambridge, Mass}, 2003.

\bibitem{TemporalDerivativeDistribution-Fishburn-2019}
F.~A. Fishburn, R.~S. Ludlum, C.~J. Vaidya, A.~V. Medvedev, Temporal
  {{Derivative Distribution Repair}} ({{TDDR}}): {{A}} motion correction method
  for {{fNIRS}}, NeuroImage 184 (2019) 171--179.
\newblock \href {https://doi.org/10.1016/j.neuroimage.2018.09.025}
  {\path{doi:10.1016/j.neuroimage.2018.09.025}}.

\bibitem{TimeSeriesMixed-Li-2014}
T.-H. Li, Time Series with Mixed Spectra, {CRC Press/Chapman \& Hall}, {Boca
  Raton, Fla.}, 2014.

\bibitem{TinnitusAltersResting-SanJuan-2017}
J.~San~Juan, X.-S. Hu, M.~Issa, S.~Bisconti, I.~Kovelman, P.~Kileny, G.~Basura,
  Tinnitus alters resting state functional connectivity ({{RSFC}}) in human
  auditory and non-auditory brain regions as measured by functional
  near-infrared spectroscopy ({{fNIRS}}), PLOS ONE 12~(6) (2017) e0179150.
\newblock \href {https://doi.org/10.1371/journal.pone.0179150}
  {\path{doi:10.1371/journal.pone.0179150}}.

\bibitem{RvwInfrared-Jue-2013}
T.~Jue, K.~Masuda (Eds.), Application of {{Near Infrared Spectroscopy}} in
  {{Biomedicine}}, {Springer US}, 2013.
\newblock \href {https://doi.org/10.1007/978-1-4614-6252-1}
  {\path{doi:10.1007/978-1-4614-6252-1}}.

\bibitem{OpticalBlood-Schmitt-1986}
J.~M. Schmitt, \href{https://books.google.no/books?id=jIgPIQAACAAJ}{Optical
  measurement of blood oxygenation by implantable telemetry}, Tech. rep.,
  {Stanford University} (1986).
\newline\urlprefix\url{https://books.google.no/books?id=jIgPIQAACAAJ}

\bibitem{MultipleScatteringField-Moaveni-1970}
M.~K. Moaveni, A {{Multiple Scattering Field Theory Applied}} to {{Whole
  Blood}}, {{PhD Thesis}}, Department of Electrical Engineering, University of
  Washington (1970).

\bibitem{CranialOptical-Duncan-1996}
A.~Duncan, J.~H. Meek, M.~Clemence, C.~E. Elwell, P.~Fallon, L.~Tyszczuk,
  M.~Cope, D.~T. Delpy, Measurement of {{Cranial Optical Path Length}} as a
  {{Function}} of {{Age Using Phase Resolved Near Infrared Spectroscopy}},
  Pediatric Research 39~(5) (1996) 889--894.
\newblock \href {https://doi.org/10.1203/00006450-199605000-00025}
  {\path{doi:10.1203/00006450-199605000-00025}}.

\bibitem{FNIR-Kassab-2015}
A.~Kassab, J.~L. Lan, P.~Vannasing, M.~Sawan, Functional near-infrared
  spectroscopy caps for brain activity monitoring: A review, Applied Optics
  54~(3) (2015) 576.
\newblock \href {https://doi.org/10.1364/AO.54.000576}
  {\path{doi:10.1364/AO.54.000576}}.

\bibitem{ImportanceNormalityAssumption-Lumley-2002}
T.~Lumley, P.~Diehr, S.~Emerson, L.~Chen, The {{Importance}} of the {{Normality
  Assumption}} in {{Large Public Health Data Sets}}, Annual Review of Public
  Health 23~(1) (2002) 151--169.
\newblock \href {https://doi.org/10.1146/annurev.publhealth.23.100901.140546}
  {\path{doi:10.1146/annurev.publhealth.23.100901.140546}}.

\bibitem{HypothesisSettingOrder-Song-2014}
C.~Song, G.~C. Tseng, Hypothesis setting and order statistic for robust genomic
  meta-analysis, The Annals of Applied Statistics 8~(2) (2014) 777--800.
\newblock \href {https://doi.org/10.1214/13-aoas683}
  {\path{doi:10.1214/13-aoas683}}.

\bibitem{StatisticalConsiderationPsy-Wilkinson-1951}
B.~Wilkinson, A statistical consideration in psychological research,
  Psychological Bulletin 48~(3) (1951) 156--158.
\newblock \href {https://doi.org/10.1037/h0059111}
  {\path{doi:10.1037/h0059111}}.

\bibitem{Vasomotion-Aalkjaer-2011}
C.~Aalkjaer, D.~Boedtkjer, V.~Matchkov, Vasomotion - what is currently
  thought?, Acta Physiologica 202~(3) (2011) 253--269.
\newblock \href {https://doi.org/10.1111/j.1748-1716.2011.02320.x}
  {\path{doi:10.1111/j.1748-1716.2011.02320.x}}.

\bibitem{SpontaneousLowFreq-Obrig-2000}
H.~Obrig, M.~Neufang, R.~Wenzel, M.~Kohl, J.~Steinbrink, K.~Einh{\"a}upl,
  A.~Villringer, Spontaneous {{Low Frequency Oscillations}} of {{Cerebral
  Hemodynamics}} and {{Metabolism}} in {{Human Adults}}, NeuroImage 12~(6)
  (2000) 623--639.
\newblock \href {https://doi.org/10.1006/nimg.2000.0657}
  {\path{doi:10.1006/nimg.2000.0657}}.

\bibitem{PulsatileTinnitus-Sila-1987}
C.~A. Sila, A.~J. Furlan, J.~R. Little, Pulsatile tinnitus., Stroke 18~(1)
  (1987) 252--256.
\newblock \href {https://doi.org/10.1161/01.STR.18.1.252}
  {\path{doi:10.1161/01.STR.18.1.252}}.

\end{thebibliography}

\end{appendices}

\end{document}